\documentclass{elsarticle}
\usepackage{hyperref}
\usepackage{amsmath}
\usepackage{amsfonts}
\usepackage{amssymb}
\usepackage{xcolor}
\usepackage{changes}

\usepackage[english]{babel}
\usepackage{lmodern}
\usepackage[T1]{fontenc}
\usepackage{units}
\usepackage{grffile}
\usepackage[bottom]{footmisc}
\usepackage{booktabs}
\usepackage{dcolumn}
\usepackage{lipsum}

\usepackage{bm}
\usepackage{graphicx}
\usepackage{epstopdf}

\DeclareMathOperator{\Tr}{Tr}

\newcounter{bla}

\journal{Computer Physics Communications}

\begin{document}
\mathchardef\mhyphen="2D

\begin{frontmatter}


  

  \title{ComDMFT v.2.0: Fully Self-Consistent \textit{ab initio} GW+EDMFT for the Electronic Structure of Correlated Quantum Materials}
  
  \author[a,b]{Byungkyun Kang}
  \author[a]{Patrick Semon}
  \author[a]{Corey Melnick}
  \author[d]{Mancheon Han}
  \author[d,e]{Seongjun Mo}
  \author[e]{Hoonkyung Lee}
  \author[a,c]{Gabriel Kotliar}
  \author[a,d]{Sangkook Choi\corref{author}}
    \cortext[author] {Corresponding author.\\\textit{sangkookchoi@kias.re.kr} S. Choi}
  \address[a]{Condensed Matter Physics and Materials Science Department,
    Brookhaven National Laboratory, Upton, NY 11973, USA}
  \address[b]{College of Arts and Sciences, University of Delaware, Newark, DE 19716, USA}
  \address[c]{Department of Physics and Astronomy, Rutgers University, NJ 08854, USA}
  \address[d]{School of Computational Sciences, Korea Institute for Advanced Study, Seoul 02455, Republic of Korea}
  \address[e]{Department of Physics, Konkuk University,Seoul 05029, Korea}
  \begin{abstract}
    
    ComDMFT is a parallel computational package designed to study the electronic structure of correlated quantum materials \textit{from first principles}. Our approach is based on the combination of \textit{first-principles} methods and dynamical mean field theories. In version 2.0, we implemented fully-diagrammatic GW+EDMFT \textit{from first-principles} self-consistently. In this approach, correlated electrons are treated within full GW+EDMFT and the rest are treated within full-GW, seamlessly. This implementation enables the electronic structure calculation of quantum materials with weak, intermediate, and strong electron correlation without prior knowledge of the degree of electron correlation.
  
\end{abstract}


\end{frontmatter}




{\bf PROGRAM SUMMARY/NEW VERSION PROGRAM SUMMARY}

\begin{small}
\noindent
{\em Program Title: ComDMFT}                                          \\
{\em Licensing provisions(please choose one): GPLv3}\\
{\em Programming language: fortran2008, C++, and Python3}                                   \\


{\em Nature of problem:}\\
In order to comprehend the electronic structure of correlated quantum materials, a non-perturbative and parameter-free \textit{ab initio} method is required. One of the promising approaches is fully-diagrammatic GW+EDMFT; however, there is no available open-source package which implements the \textit{ab initio} full GW+EDMFT methodology.
\\

{\em Solution method:}\\
We implemented \textit{ab initio} all electron full GW+EDMFT methodology for the electronic structure calculation of correlated quantum materials  and compare the results with those of LQSGW+DMFT.\\

{\em Additional comments:}\\
ComDMFTv2 is built on top open source codes,
ComDMFTv1 \cite{Choi_Kotliar-ComDMFTMassively-ComputerPhysicsCommunications-2019},
Wannier90 \cite{Mostofi_Marzari-Wannier90Tool-Comput.Phys.Commun.-2008},  ComCTQMC\cite{Melnick_Kotliar-AcceleratedImpurity-ComputerPhysicsCommunications-2021} and LqsgwFlapw \cite{Kutepov_Kotliar-LinearizedSelfconsistent-ComputerPhysicsCommunications-2017} codes. ComDMFTv2 underwent testing on Intel supercomputers with intel compilers and future revisions will ensure compatibility with other compilers.\\

\end{small}

\section{Introduction}

In materials with a large number of electrons, the quantum mechanical interactions between these electrons often dictate the materials properties. These so-called correlated quantum materials can display a range of exceptional emergent phenomena, including metal-insulator transition, colossal magnetoresistance, and high-temperature superconductivity. Recently, correlated quantum materials received renewed attentions due to their potential for quantum information science applications. An excellent example is the Fe-based superconductors, which have demonstrated potential as qubit hosts which are topologically-protected. \cite{Xu_Zhang-TopologicalSuperconductivity-Phys.Rev.Lett.-2016,Wang_Gao-EvidenceMajorana-Science-2018, Zhang_Shin-ObservationTopological-Science-2018,Fernandes_Kotliar-IronPnictides-Nature-2022}.

One of the grand challenges in condensed matter physics as well as materials science is to predict correlated quantum materials properties \textit{from first principles}. 
Treating correlated quantum materials require new concepts, algorithms, as well as computer codes. To this end, various ideas have been realized and applied for the quantitative understanding of correlated quantum materials properties \cite{Kent_Kotliar-PredictiveTheory-Science-2018}. Among them, one of the most successful methodologies is dynamical mean-field theory (DMFT) \cite{Metzner_Vollhardt-CorrelatedLattice-Phys.Rev.Lett.-1989,Muller-Hartmann_Muller-Hartmann-CorrelatedFermions-Z.PhysikB-CondensedMatter-1989,Brandt_Mielsch-ThermodynamicsCorrelation-Z.PhysikB-CondensedMatter-1989,Janis_Janis-NewConstruction-Z.PhysikB-CondensedMatter-1991,Georges_Kotliar-HubbardModel-Phys.Rev.B-1992,Jarrell_Jarrell-HubbardModel-Phys.Rev.Lett.-1992,Rozenberg_Kotliar-MottHubbardTransition-Phys.Rev.Lett.-1992,Georges_Krauth-NumericalSolution-Phys.Rev.Lett.-1992,Georges_Rozenberg-DynamicalMeanfield-Rev.Mod.Phys.-1996,Adler_Kotliar-CorrelatedMaterials-Rep.Prog.Phys.-2018}. DMFT establishes a mapping from a quantum many-body problem onto a quantum impurity problem. Quantities obtained within quantum impurity problems are embedded into the Green's function of the quantum many-body problems. This idea has been applied extensively to model Hamiltonians 
and has described correlated electron phenomena  such as  Mott  physics,  \cite{Rozenberg_Kotliar-MottHubbardTransition-Phys.Rev.Lett.-1992,Zhang_Kotliar-MottTransition-Phys.Rev.Lett.-1993}  Hund metal physics \cite{deMedici_Georges-JanusFacedInfluence-Phys.Rev.Lett.-2011} and heavy fermion physics \cite{Jang_Shim-EvolutionKondo-PNAS-2020}.

The success of model-based DMFT approaches has led to the integration of DMFT with \textit{ab initio} methodologies. This combination is based on the concept of quantum embedding, which entails separating electrons into two groups \cite{Anisimov_Andersen-BandTheory-Phys.Rev.B-1991,Anisimov_Sawatzky-DensityfunctionalTheory-Phys.Rev.B-1993,Liechtenstein_Zaanen-DensityfunctionalTheory-Phys.Rev.B-1995}: correlated electrons and itinerant electrons. Correlated electron phenomena are addressed by DMFT, while \textit{first-principles} methods are used to describe itinerant electrons and to generate materials-specific quantum impurity problems. An example of this successful union is the DFT+DMFT methodology \cite{Anisimov_Kotliar-FirstprinciplesCalculations-J.Phys.:Condens.Matter-1997,Lichtenstein_Katsnelson-InitioCalculations-Phys.Rev.B-1998,Kotliar_Marianetti-ElectronicStructure-Rev.Mod.Phys.-2006}. This approach can explain various phenomena specific to correlated quantum materials such as paramagnetic Mott transitions\cite{Held_Anisimov-MottHubbardMetalInsulator-Phys.Rev.Lett.-2001} and volume collapse transitions\cite{Savrasov_Abrahams-CorrelatedElectrons-Nature-2001}. The development of the continuous-time Monte-Carlo approaches to quantum impurity problems \cite{Werner_Millis-ContinuousTimeSolver-Phys.Rev.Lett.-2006} has enabled the analysis of broad classes of quantum impurity problems, making them the popular choice for DMFT-based \textit{ab initio} calculations.  Recent important advances in the DFT+DMFT  methodology were introduced to compute total energies\cite{Park_Marianetti-ComputingTotal-Phys.Rev.B-2014}, free energies\cite{Haule_Birol-FreeEnergy-Phys.Rev.Lett.-2015}, and forces\cite{Haule_Pascut-ForcesStructural-Phys.Rev.B-2016}, as well as more accurate  double-counting scheme corrections\cite{Haule_Haule-ExactDouble-Phys.Rev.Lett.-2015}.

The DFT+DMFT methodology has attracted a lot of attention  and is supported by the development of various software programs. Reusable libraries, such as TRIQS\cite{Parcollet_Seth-TRIQSToolbox-ComputerPhysicsCommunications-2015} and ALPS\cite{Gaenko_Gull-UpdatedCore-ComputerPhysicsCommunications-2017}, have been created to aid DFT+DMFT implementation. Standalone impurity solvers, iQIST\cite{Huang_Dai-IQISTOpen-ComputerPhysicsCommunications-2015}, W2dynamics\cite{Wallerberger_Sangiovanni-W2dynamicsLocal-ComputerPhysicsCommunications-2019} and ComCTQMC\cite{Melnick_Kotliar-AcceleratedImpurity-ComputerPhysicsCommunications-2021}, are available. DFT+DMFT packages
have been developed: ComDMFT\cite{Choi_Kotliar-ComDMFTMassively-ComputerPhysicsCommunications-2019}; EDMFTF\cite{Haule_Kim-DynamicalMeanfield-Phys.Rev.B-2010} integrated with Wien2K\cite{Blaha_Marks-WIEN2kAPW-J.Chem.Phys.-2020}; TRIQS/DFTTools \cite{Aichhorn_Parcollet-TRIQSDFTTools-ComputerPhysicsCommunications-2016} built on top of TRIQS\cite{Parcollet_Seth-TRIQSToolbox-ComputerPhysicsCommunications-2015} and Wien2k\cite{Blaha_Marks-WIEN2kAPW-J.Chem.Phys.-2020}; DMFTwDFT\cite{Singh_Park-DMFTwDFTOpensource-ComputerPhysicsCommunications-2021,Park_Marianetti-ComputingTotal-Phys.Rev.B-2014} bulit on top of EDMFTF package impurity solver\cite{Haule_Haule-QuantumMonte-Phys.Rev.B-2007}; D-Core \cite{_-DCoreDCore--} built on top of TRIQS\cite{Aichhorn_Parcollet-TRIQSDFTTools-ComputerPhysicsCommunications-2016,Parcollet_Seth-TRIQSToolbox-ComputerPhysicsCommunications-2015}, ALPS\cite{Gaenko_Gull-UpdatedCore-ComputerPhysicsCommunications-2017}, Quantum Espresso\cite{Giannozzi_Wentzcovitch-QUANTUMESPRESSO-J.Phys.:Condens.Matter-2009}, and OpenMX\cite{_-OpenMXWebsite--}; AMULET\cite{_-Amulet--} integrated with Quantum Espresso\cite{Giannozzi_Wentzcovitch-QUANTUMESPRESSO-J.Phys.:Condens.Matter-2009} and Elk\cite{_-ElkCode--}; and Wien2k+W2dynamics. Furthermore, DFT+DMFT functionality has been added to \textit{ab initio} codes such as Abinit\cite{Gonze_Zwanziger-RecentDevelopments-ComputerPhysicsCommunications-2016} and RSPt\cite{Wills_Grechnyev-FullPotentialElectronic-Full-PotentialElectronicStructureMethod-2010}.

Similarly, the rotationally invariant slave bosons (RISB), which is equivalent to the Gutzwiller method,  were combined with density functional theory in the LDA+RISB or LDA+G method \cite{Deng_Fang-LocalDensity-Phys.Rev.B-2009,Yao_Ho-IncludingManybody-Phys.Rev.B-2011,Ho_Wang-GutzwillerDensity-Phys.Rev.B-2008,Schickling_Weber-GutzwillerDensity-NewJ.Phys.-2014}. They were also cast as a quantum embedding method \cite{Lanata_Kotliar-PhaseDiagram-Phys.Rev.X-2015a,Lanata_Kotliar-SlaveBoson-Phys.Rev.Lett.-2017a} and open source implementations of this approach such as CyGutz and ComGutz  are available \cite{_-WelcomeCyGutz--,_-DownloadsComscope--}. More recently,  direct comparisons of LDA+DMFT and LDA+G, including total energies,  have been carried out  using the Portobello framework \cite{Adler_Kotliar-PortobelloQuantum-ComputerPhysicsCommunications-2024}. These approaches can also be combined with GW methods for the weakly correlated subsystems.

Density Functional Theory combined with Dynamical Mean Field Theory (DFT+DMFT) has been very successful in predicting the properties of static equilibrium properties of correlated quantum materials. Moreover, it has been proven that the predicted dynamics of electronic excitations agrees well with experimental data from many spectroscopic tools.  It requires, however, the introduction of parameters in constructing quantum impurity problem. For examples, Coulomb interactions of the correlated subshell are required.
For DFT+DMFT, the Coulomb interaction can be obtained by using  cRPA~\cite{Aryasetiawan_Lichtenstein-FrequencydependentLocal-Phys.Rev.B-2004,Aryasetiawan_Schonberger-CalculationsHubbard-Phys.Rev.B-2006} as well as constrained DFT~\cite{anisimov2009coulomb}.

The success of DFT+DMFT has motivate the search for its justification and extension via diagrammatic Green's function approaches based on free energy functionals \cite{Baym_Kadanoff-ConservationLaws-Phys.Rev.-1961,Luttinger_Ward-GroundStateEnergy-Phys.Rev.-1960,Potthoff_Potthoff-SelfenergyfunctionalApproach-Eur.Phys.J.B-2003}. By defining exchange-correlational functionals using Green's function ($G$) and the screened Coulomb interaction ($W$)\cite{Almbladh_Leeuwen-VariationalTotal-Int.J.Mod.Phys.B-1999,Chitra_Kotliar-EffectiveactionApproach-Phys.Rev.B-2001}, these approaches provide a systematic way to calculate the properties of materials properties. Diagrammatic approaches have proven successful in \textit{first-principles} implementations. One good example is the GW approximation for studying itinerant electron systems. This method employs the first-order diagram in $W$ to represent the electron-boson coupling in the free-energy functional \cite{Almbladh_Leeuwen-VariationalTotal-Int.J.Mod.Phys.B-1999}, which can be interpreted as an approximation of the bare three-point vertex\cite{Hedin_Hedin-NewMethod-Phys.Rev.-1965}. The use of this \textit{ab initio} method \cite{Strinati_Hanke-DynamicalCorrelation-Phys.Rev.Lett.-1980,Strinati_Hanke-DynamicalAspects-Phys.Rev.B-1982, Hybertsen_Louie-FirstPrinciplesTheory-Phys.Rev.Lett.-1985,Hybertsen_Louie-ElectronCorrelation-Phys.Rev.B-1986,Godby_Sham-QuasiparticleEnergies-Phys.Rev.B-1987} has significantly improved the accuracy of quasiparticle spectra of semincoductors and insulators in comparison to DFT. The development of software programs such as ABINIT\cite{Gonze_Zwanziger-RecentDevelopments-ComputerPhysicsCommunications-2016}, BerkeleyGW\cite{Deslippe_Louie-BerkeleyGWMassively-Comput.Phys.Commun.-2012}, CP2K\cite{Kuhne_Hutter-CP2KElectronic-TheJournalofChemicalPhysics-2020}, Elk\cite{_-ElkCode--}, FHI-aims\cite{Blum_Scheffler-InitioMolecular-ComputerPhysicsCommunications-2009}, Fiesta\cite{Blase_Olevano-FirstprinciplesGW-Phys.Rev.B-2011}, LqsgwFlapw\cite{Kutepov_Kotliar-LinearizedSelfconsistent-ComputerPhysicsCommunications-2017}, GPAW\cite{Huser_Thygesen-QuasiparticleGW-Phys.Rev.B-2013}, PySCF\cite{Sun_Chan-RecentDevelopments-TheJournalofChemicalPhysics-2020}, Questaal\cite{Pashov_VanSchilfgaarde-QuestaalPackage-ComputerPhysicsCommunications-2020}, VASP\cite{_-VASPVienna--}, West\cite{Govoni_Galli-LargeScale-J.Chem.TheoryComput.-2015a}, and Yambo\cite{Sangalli_Marini-ManybodyPerturbation-J.Phys.:Condens.Matter-2019} has facilitated the usage of ab initio GW methodology.

For the correlated quantum materials, a combination of many-body perturbation theory (MBPT) and DMFT can be used through this functional approach. The simplest example of this functional approach is the GW+EDMFT approach \cite{Sun_Kotliar-ExtendedDynamical-Phys.Rev.B-2002,Biermann_Georges-FirstPrinciplesApproach-Phys.Rev.Lett.-2003,Nilsson_Aryasetiawan-MultitierSelfconsistent-Phys.Rev.Materials-2017}; by accounting for exchange and correlation effects within the GW approximation for itinerant electrons and extended DMFT\cite{Si_Smith-KosterlitzThoulessTransition-Phys.Rev.Lett.-1996,Sengupta_Georges-NonFermiliquidBehavior-Phys.Rev.B-1995,HenrikKajueter_HenrikKajueter-InterpolatingPerturbation--1996} for correlated electrons, this approach merges these two theories. When GW methodology is merged with extended DMFT, this approach is advantageous as it is free of adjustable parameters; once  a projector is chosen to  describe the correlated orbitals, the method proceeds without any further input from the user and defines the quantum impurity problem for the correlated orbitals self-consistently. Similar approaches with the same goal include \textit{ab initio} Density Matrix Embedding Theory (DMET) \cite{Cui_Chan-EfficientImplementation-J.Chem.TheoryComput.-2020}, full-cell DMFT \cite{Zhu_Chan-InitioFull-Phys.Rev.X-2021}, and Self-Energy Embedding Theory (SEET) \cite{Lan_Zgid-TestingSelfenergy-Phys.Rev.B-2017}. Originally, full-cell DMFT and SEET were formulated using the Green’s function ($G$) and the bare Coulomb interaction ($V$), but they have recently been extended to incorporate the screened Coulomb interaction ($W$).


 Various earlier schemes of GW+EDMFT have been proposed. Two examples are the partial self-consistent scheme \cite{Tomczak_Tomczak-QSGWDMFT-J.Phys.:Conf.Ser.-2015} and the multitier GW+EDMFT approach\cite{Nilsson_Aryasetiawan-MultitierSelfconsistent-Phys.Rev.Materials-2017}. These combine GW and DMFT, but not in a fully self consistent manner.  The former integrates DMFT with quasiparticle GW approaches such as one-shot GW\cite{Tomczak_Biermann-AsymmetryBand-Phys.Rev.B-2014,Taranto_Held-ComparingQuasiparticle-Phys.Rev.B-2013,Werner_Biermann-SatellitesLarge-NatPhys-2012}, QSGW\cite{Sponza_Kotliar-SelfenergiesItinerant-Phys.Rev.B-2017,Pashov_VanSchilfgaarde-QuestaalPackage-ComputerPhysicsCommunications-2020}, LQSGW\cite{Choi_Kotliar-FirstprinciplesTreatment-NpjQuantumMater.-2016,Choi_Kotliar-ComDMFTMassively-ComputerPhysicsCommunications-2019}, and screened exchange\cite{vanRoekeghem_Biermann-DynamicalCorrelations-Phys.Rev.Lett.-2014}. Coulomb interactions of the correlated subshells are obtained within constrained random phase approximation\cite{Aryasetiawan_Lichtenstein-FrequencydependentLocal-Phys.Rev.B-2004}. In these approaches, DMFT is treated as a one-shot correction to the corresponding quasiparticle Hamiltonian. On the other hand, the multitier GW+EDMFT entails self-consistency within a subspace of the full Hilbert space, while the remaining Hilbert space is treated with a one-shot GW. These approaches have been applied to various correlated quantum materials \cite{Petocchi_Werner-FullyInitio-Phys.Rev.B-2021,Petocchi_Werner-NormalState-Phys.Rev.X-2020,Petocchi_Werner-Screeninge-Phys.Rev.Research-2020,Boehnke_Werner-WhenStrong-Phys.Rev.B-2016}, exhibiting the capability of full implementation. Within the field of correlated quantum materials, there is a pressing requirement for a package that implements the \textit{ab initio} full GW+EDMFT methodology. Currently, few such packages exist and none of them  supports fully self-consistent GW+DMFT scheme. ComDMFT\cite{Choi_Kotliar-ComDMFTMassively-ComputerPhysicsCommunications-2019}  supports the \textit{ab initio} LQSGW+DMFT methodology \cite{Choi_Kotliar-FirstprinciplesTreatment-NpjQuantumMater.-2016} and Questaal\cite{Pashov_VanSchilfgaarde-QuestaalPackage-ComputerPhysicsCommunications-2020} implements the QSGW+DMFT methodology. 

This paper describes the ComDMFT v2.0 package, developed at Brookhaven National Lab. It is the first implementation of full GW+EDMFT.  As an open source code for the community, it builds on  earlier open source packages  ComDMFT v.1.0 \cite{Choi_Kotliar-ComDMFTMassively-ComputerPhysicsCommunications-2019}, the LqsgwFlapw code\cite{Kutepov_Kotliar-LinearizedSelfconsistent-ComputerPhysicsCommunications-2017} for \textit{ab initio} GW calculations, the Wannier90 code\cite{Mostofi_Marzari-Wannier90Tool-Comput.Phys.Commun.-2008} for the construction of correlated orbitals, and the ComCTQMC code\cite{Melnick_Kotliar-AcceleratedImpurity-ComputerPhysicsCommunications-2021} for the solution of quantum impurity problems. The package is composed of five modules, written in FORTRAN2008 and Python3, namely: ComBasis, Comlocal, ComWeiss, ComDC, and ComEmbed. 

In this paper, we introduce full GW+EDMFT methodology and to provide benchmark results on SrVO$_3$ and NiO. Section \ref{sec_gen_theo_framework} outlines the general theoretical framework, while Section \ref{sec_com_layout} provides a detailed description of our implementation. Benchmark results are presented in Section \ref{sec_results}  together with  calculations with LQSGW+DMFT for the same materials for comparison. 

\section{Theoretical Background} \label{sec_gen_theo_framework}

A very useful approach in many body theory is  provided by  the functional formalism of Luttinger and  Ward \cite{Luttinger_Ward-GroundStateEnergy-Phys.Rev.-1960} and Baym and Kadanoff \cite{Baym_Kadanoff-ConservationLaws-Phys.Rev.-1961}. Here, the grand potential ($\Omega$) is  viewed as a functional of the electron Green's function ($G$) and electron self-energy ($\Sigma$). $\Omega[G, \Sigma]$ is stationary at the physical $G$ and the corresponding $\Sigma$.  Approximations to this exact functional result in approximate many body methods. 
This approach has been later extended to construct a new functional, $\Gamma$:
\cite{Almbladh_Leeuwen-VariationalTotal-Int.J.Mod.Phys.B-1999,Chitra_Kotliar-EffectiveactionApproach-Phys.Rev.B-2001}

\begin{equation}
  \begin{split}
    \Gamma[G,\Sigma,W,\Pi]=&-\Tr[log(-G_H^{-1})]-\Tr[log(1-G_H\Sigma)]-\Tr[G\Sigma]\\
                           &+\frac{1}{2}\Tr[log(1-V\Pi)]+\frac{1}{2}\Tr[\Pi W]+\Psi[G,W].\\
  \end{split}
\end{equation}
Here, $\Tr$ is used to denote the sum over all relevant indices (crystal momentum, state indices, and spin indices) in addition to Matsubara frequency summation, with a prefactor of $1/\beta$. $G_H$ and $V$ denote the Hartree Green's function and bare Coulomb interaction, respectively.
$\Psi$ is a functional expressed as a summation of all two-particle irreducible diagrams constructed from $G$ and $W$. Approximations to $\Psi$ result in approximate many body methods. Within this construction, $\Gamma$ is regarded as the functional of Green's function ($G$), electron self-energy ($\Sigma$), screened Coulomb interaction($W$), and polarizability ($\Pi$). These can be shown to be stationary at the physical $G$ and $W$, along with the corresponding $\Sigma$ and $\Pi$. These conditions lead to the following equations:

\begin{equation}
  \begin{split}
    \frac{\delta \Gamma}{\delta G}=0\to& \Sigma=\frac{\delta \Psi}{\delta G}\\
    \label{eq:f_selfenergy}            
  \end{split}
\end{equation}
\begin{equation}
  \begin{split}
    \frac{\delta \Gamma}{\delta \Sigma}=0\to& G=\left(G_H^{-1}-\Sigma\right)^{-1}\\
    \label{eq:f_dyson}                
  \end{split}
\end{equation}
\begin{equation}
  \begin{split}
    \frac{\delta \Gamma}{\delta W}=0\to& \Pi=-2\frac{\delta \Psi}{\delta W}\\
    \label{eq:b_selfenergy}                
  \end{split}
\end{equation}
\begin{equation}
  \begin{split}
    \frac{\delta \Gamma}{\delta \Pi}=0\to& W=\left(V^{-1}-\Pi\right)^{-1}\\
    \label{eq:b_dyson}        
  \end{split}
\end{equation}

This approach can be utilized to comprehend various diagrammatic approximations. For instance, the first-order diagram in $W$ ($\Psi_{GW}(G,W)= -\frac{1}{2}\Tr GWG$) corresponds to the renowned GW approximation\cite{Hedin_Hedin-NewMethod-Phys.Rev.-1965,Almbladh_Leeuwen-VariationalTotal-Int.J.Mod.Phys.B-1999}. $\Psi_{GW}(G,W)$ leads to the following self-energy and polarizability.

\begin{equation}
  \begin{split}
    \Sigma_{GW}(1,2)&=\frac{\delta \Psi_{GW}}{\delta G(2,1)}=-G(1,2)W(1,2)\\
    \Pi_{GW}(1,2)&=-2\frac{\delta \Psi_{GW}}{\delta W(2,1)}=G(1,2)G(2,1)\\
    \label{eq:gw_fb_self}        
  \end{split}
\end{equation}
Here, the notation $1$ denotes $(\mathbf{r}_1,\sigma_1,\tau_1)=(\mathbf{x}_1, \tau_1)=x_1$, where $\mathbf{r}$, $\sigma$ and $\tau$ are the position vector, spin index, and imaginary time, respectively. 

In order to incorporate strong correlation effects, the DMFT idea can be applied. This amounts to constraining the $G$ and $W$ in the $\Psi$ functional to the correlated orbital space, and is referred to as extended DMFT (EDMFT)\cite{Si_Smith-KosterlitzThoulessTransition-Phys.Rev.Lett.-1996,Sengupta_Georges-NonFermiliquidBehavior-Phys.Rev.B-1995,HenrikKajueter_HenrikKajueter-InterpolatingPerturbation--1996}. Rigorously, the correlated orbitals $\varphi_\alpha^f(\mathbf{r})$ are introduced, where $\alpha$ is a composite index for the correlated orbital and spin, and $f$ denotes ``fermion''. The projection ($\hat{P}^f$) and embedding ($\hat{E}^f$) of the fermionic quantities can then be defined accordingly.

\begin{equation}
  \begin{split}
    \widetilde {G}_{\alpha\beta}(\tau_1,\tau_2)=&\hat{P}_{\alpha\beta}^{f}(G)(\tau_1,\tau_2)=\int d\mathbf{x}_1\mathbf{x}_2 {\varphi_{\alpha}^f}^*(\mathbf{x}_1)G(\mathbf{x}_1,\tau_1,\mathbf{x}_2,\tau_2)  \varphi_{\beta}^f(\mathbf{x}_2)\\
    \widetilde{G}(1,2)=&\hat{E}^{f}(\widetilde{G})(1,2)=\sum_{\alpha\beta} \varphi_{\alpha}^f(\mathbf{x}_1)\widetilde{G}_{\alpha\beta}(\tau_1,\tau_2)  {\varphi_{\beta}^f}^{*}(\mathbf{x}_2)
  \end{split}
  \label{eq:f_projection_embedding}              
\end{equation}
Here, $\ \widetilde{}\ $ denotes a physical quantity projected onto the correlated orbital space and $\int d\mathbf{x}_1=\sum_{\sigma_1}\int d\mathbf{r}_1$.

For the projection ($\hat{P}^b$) and embedding ($\hat{E}^b$) of bosonic quantities, it is convenient to define orthonormalized product basis functions. To develop these functions, we initially construct unorthonormalized product basis functions, $\vartheta$. 
\begin{equation}
  \begin{split}
    \vartheta_{\alpha\beta}(\mathbf{x})={\varphi_{\alpha}^f}^*(\mathbf{x}){\varphi_{\beta}^f}(\mathbf{x})
  \end{split}
  \label{eq:f_projection}              
\end{equation}
The orthonormalized product basis function, $\varphi^b$ can be represented as a linear combination of $\vartheta$:
\begin{equation}
  \begin{split}
    \varphi_{\alpha}^b(\mathbf{x})=\sum_{\beta\gamma}\vartheta_{\beta\gamma}(\mathbf{x})X_{\beta\gamma,\alpha}
  \end{split}
\end{equation}
For the construction of $\varphi^{b}$, we employed canonical orthonormalization. The overlap matrix ($S_{\alpha\beta;\gamma\delta}=\int d\mathbf{x} \vartheta_{\alpha\beta}^*(\mathbf{x}) \vartheta_{\gamma\delta}(\mathbf{x})$) was diagonalized by a unitary matrix $U$.
\begin{equation}
  \begin{split}
    U^{\dagger}SU=D
  \end{split}
\end{equation}
where $D$ is the diagonal matrix of the eigenvalues of $S$. Then $X$ is given by
\begin{equation}
  \begin{split}
    X_{\beta\gamma,\alpha}=U_{\beta\gamma;\alpha}D_\alpha^{-1/2}
  \end{split}
\end{equation}
By using the orthonormalized product basis, we can define the projection and embedding of the bosonic quantities in the following way.
\begin{equation}
  \begin{split}
    \widetilde {W}_{\alpha\beta}(\tau_1,\tau_2)=&\hat{P}_{\alpha\beta}^{b}(W)(\tau_1,\tau_2)=\int d\mathbf{x}_1d\mathbf{x}_2 {\varphi_{\alpha}^b}^*(\mathbf{x}_1)W(\mathbf{x}_1,\tau_1,\mathbf{x}_2,\tau_2)  {\varphi_{\beta}^b}(\mathbf{x}_2)\\
    \widetilde{W}(1,2)=&\hat{E}^{b}(\widetilde{W})(1,2)=\sum_{\alpha\beta} \varphi_{\alpha}^b(\mathbf{x}_1)\widetilde{W}_{\alpha\beta} (\tau_1,\tau_2) {\varphi_{\beta}^b}^{*}(\mathbf{x}_2)  \end{split}
  \label{eq:b_projection_embedding}              
\end{equation}
Alternatively, the projected bosonic quantities can be defined by using unorthonormalized product basis.
\begin{equation}
  \begin{split}
    \widetilde {W}_{\alpha\beta\gamma\delta}=&\hat{P}^{b}_{\alpha\beta\gamma\delta}(W)(\tau_1,\tau_2)=\int d\mathbf{x}_1d\mathbf{x}_2 {\vartheta_{\alpha\delta}}^*(\mathbf{x}_1)W(\mathbf{x}_1,\tau_1,\mathbf{x}_2\tau_2)  \vartheta_{\gamma\beta}(\mathbf{x}_2)\\
    \label{eq:b_projection_fourindces}
    \end{split}    
  \end{equation}

\begin{figure}
  \centering
  \includegraphics[width=1.0\columnwidth]{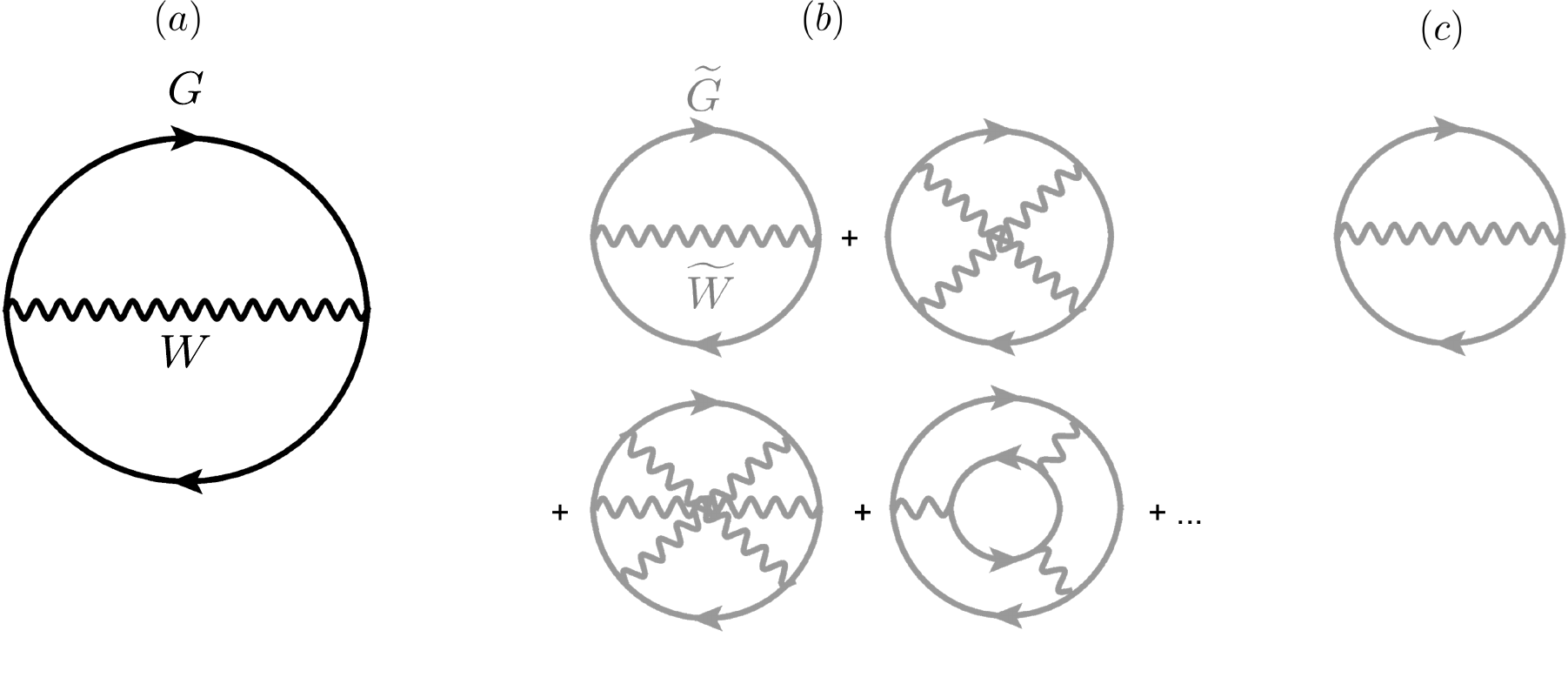}
  \caption{Diagrammatic representations of the $\Psi$ functional within full GW+EDMFT approach. (a) GW $\Psi$  functional (b) EDMFT $\Psi$ functional and (c) local-GW $\Psi$ functional. Full and wiggly lines represent Green's function ($G$) and screened Coulomb interaction ($W$), respectively. Black and gray colors represent quantities in the full Hilbert space and in the correlated subspace, respectively.}
  \label{fig_psi_functional}
\end{figure}  

  The EDMFT $\Psi$ functional ($\Psi_{EDMFT}(\widetilde{G},\widetilde{W})$) is given by all two-particle irreducible diagrams constructed from $\widetilde{G}$ and $\widetilde{W}$. The electron self-energy and polarizability within EDMFT is given by the following equations.
\begin{equation}
  \begin{split}
    \Sigma_{EDMFT}(1,2)=\frac{\delta \Psi_{EDMFT}}{\delta G(2,1)}=\frac{\delta \Psi_{EDMFT}}{\delta \widetilde{G}(2,1)}\\
    \Pi_{EDMFT}(1,2)=-2\frac{\delta \Psi_{EDMFT}}{\delta W(2,1)}=-2\frac{\delta \Psi_{EDMFT}}{\delta \widetilde{W}(2,1)}\\
  \end{split}
  \label{eq:emdft_fb_self}              
\end{equation}

The functional approach provides a natural way to merge these two diagrammatic approaches \cite{Sun_Kotliar-ExtendedDynamical-Phys.Rev.B-2002,Biermann_Georges-FirstPrinciplesApproach-Phys.Rev.Lett.-2003,Nilsson_Aryasetiawan-MultitierSelfconsistent-Phys.Rev.Materials-2017}. When we merge $\Psi_{GW}(G,W)$ and $\Psi_{EDMFT}(\widetilde{G},\widetilde{W})$, the double-counted diagrams should be subtracted. In GW+EDMFT, the GW $\Psi$ functional in the correlated space ($\Psi_{DC}(\widetilde{G},\widetilde{W})= -\frac{1}{2}\Tr \widetilde{G}\widetilde{W}\widetilde{G}$) provides the double counted diagrams.  In particular, the double-counting self-energy and polarizability are given by
\begin{equation}
  \begin{split}
    \Sigma_{DC}(1,2)&=\frac{\delta \Psi_{DC}}{\delta G(2,1)}=\frac{\delta \Psi_{DC}}{\delta \widetilde{G}(2,1)}=-\widetilde{G}(1,2)\widetilde{W}(1,2)\\
    \Pi_{DC}(1,2)&=-2\frac{\delta \Psi_{DC}}{\delta W(2,1)}=-2\frac{\delta \Psi_{DC}}{\delta \widetilde{W}(2,1)}=\widetilde{G}(1,2)\widetilde{G}(2,1)\\
    \label{eq:gw_fb_self}        
  \end{split}
\end{equation}

Finally, we can derive the $\Psi$ functional of GW+EDMFT ($\Psi_{GW+EDMFT}(G,W)$) illustrated in Fig. \ref{fig_psi_functional}.
\begin{equation}
  \begin{split}
    \Psi_{GW+EDMFT}(G,W)= -\frac{1}{2}\Tr GWG+\Psi_{EDMFT}(\widetilde{G},\widetilde{W})+\frac{1}{2}\Tr \widetilde{G}\widetilde{W}\widetilde{G}.
  \end{split}
\end{equation}
Then, the electron self-energy within GW+EDMFT is given by the following equations. For the correlated states, 
\begin{equation}
  \Sigma(1,2)=-G(1,2)W(1,2)+\frac{\delta \Psi_{EDMFT}}{\delta \widetilde{G}(2,1)}+\widetilde{G}(1,2)\widetilde{W}(1,2).
  \label{eq:selfenergy_gwedmft}                    
\end{equation}
For the itinerant states,
\begin{equation}
  \Sigma_{GW}(1,2)=-G(1,2)W(1,2).
    \label{eq:selfenergy_gw}                    
\end{equation}
The polarizability within GW+EDMFT is given by the following equations. For the correlated states,
\begin{equation}
  \Pi(1,2)=G(1,2)G(2,1)-2\frac{\delta \Psi_{EDMFT}}{\delta \widetilde{W}(2,1)}-\widetilde{G}(1,2)\widetilde{G}(2,1).
    \label{eq:pol_gwedmft}                    
  \end{equation}
For the itinerant states,
\begin{equation}
  \Pi(1,2)=G(1,2)G(2,1).
    \label{eq:pol_gw}                    
\end{equation}  

\begin{figure}
  \centering
  \includegraphics[width=1.0\columnwidth]{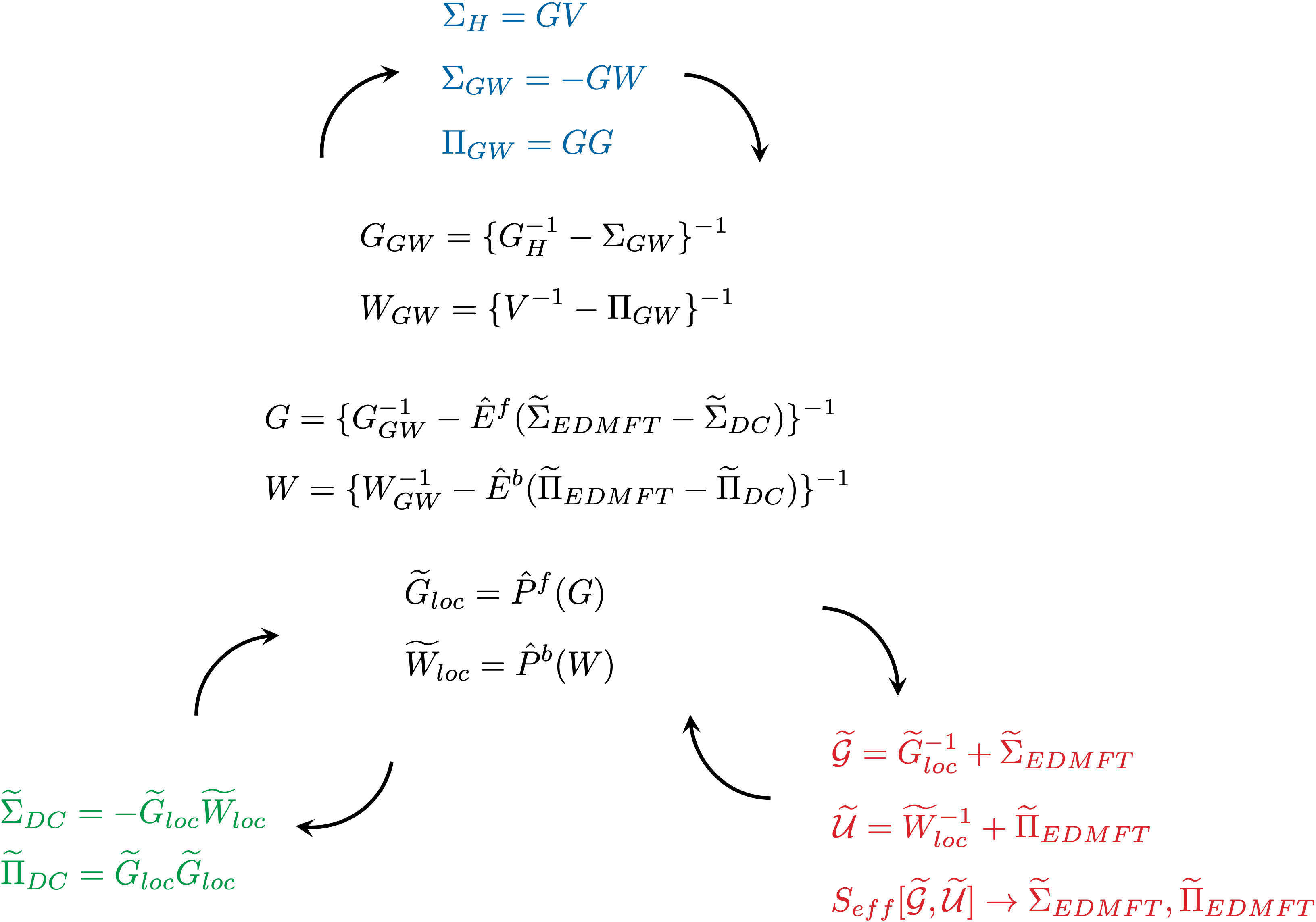}
  \caption{ The GW+EDMFT approach involves three self-consistency loops: the GW loop (blue), the EDMFT loop (red), and the double counting loop (green). This diagram seeks to emphasize the need for self-consistent determination of three-different terms in both fermionic and bosonic self-energies. For more information on the practical implementation of the self-consistent loop, please refer to Fig. \ref{flowchart}.}
  \label{fig_gwedmft_three_scf}
\end{figure}
The GW+EDMFT approximation provides a set of self-consistent equations, which can be solved to obtain $G$, $\Sigma$, $W$, and $\Pi$ at their stationary conditions. Central equations are the fermionic and bosonic Dyson's equations, as shown in eqs. (\ref{eq:f_dyson}) and (\ref{eq:b_dyson}), respectively. In addition, three distinct self-consistency loops must be considered: the GW loop, EDMFT loop, and double counting loop. A summary of these equations and loops is provided in Fig. \ref{fig_gwedmft_three_scf}. Here we note that Dyson's equations for Green's function can be written in terms of $G_{GW}$
\begin{equation}
  G_{GW}=\left(G_H^{-1}-\Sigma_{GW}\right)^{-1}
  \label{eq:g_gw}                  
\end{equation}
\begin{equation}
  G=\{G_{GW}^{-1}-\hat{E}^f(\widetilde{\Sigma}_{EDMFT}-\widetilde{\Sigma}_{DC})\}^{-1}.
  \label{eq:f_dyson_w_gw}                    
\end{equation}
In the same way, Dyson's equation for screened Coulomb interaction can be written in terms of  $W_{GW}$
\begin{equation}
  W_{GW}=\left(V^{-1}-\Pi_{GW}\right)^{-1}
  \label{eq:w_gw}                  
\end{equation}
\begin{equation}
  W=\{W_{GW}^{-1}-\hat{E}^b(\widetilde{\Pi}_{EDMFT}-\widetilde{\Pi}_{DC})\}^{-1}
  \label{eq:b_dyson_w_gw}                    
\end{equation}

\section{ \text{Ab initio} fully self-consistent  GW+EDMFT implementation in ComDMFT v2.0}\label{sec_com_layout}
ComDMFT v2.0 comprises eight components and a Python script: LqsgwFlapw\cite{Kutepov_Kotliar-LinearizedSelfconsistent-ComputerPhysicsCommunications-2017}, Wannier90\cite{Mostofi_Marzari-Wannier90Tool-Comput.Phys.Commun.-2008}, ComBasis, ComWeiss, ComLocal, ComCTQMC\cite{Melnick_Kotliar-AcceleratedImpurity-ComputerPhysicsCommunications-2021}, ComDC, and ComEmbed. These components are executed in sequence by the Python script ``gwdmft.py''. 

All the other components are developed using object-oriented programming features in python3 and fortran 2008. The implemented resources (classes) are reused by several components and can easily be reused by future developers. All the generated data by individual components are saved in a single hdf5 file and can be accessed by other components based on a data-map upon inquiry with keys (file name, group name, data name).


For the GW+EDMFT implementation, we will consider three Hilbert spaces: the full space (F), the correlated space (C), and the low-energy space (L). The full space is the entire Hilbert space. To label the basis functions in the full space, we will use an upper-case Latin alphabet $|\varphi_I^{f/b}\rangle$ or a position vector together with a spin index $|\mathbf{x}\rangle=|\mathbf{r}\sigma\rangle$. Here $f$ and $b$ in the superscript denote ``Fermion'' and ``Boson'' respectively. The correlated space is a subspace of the full space. They are spanned by correlated orbitals for fermionic quantities and orthonormalized product basis sets of the correlated orbitals for bosonic quantities. We will use a lower-case Greek alphabet to label the basis functions in the correlated space, e.g. $|\varphi_\alpha^{f/b}\rangle$.

\begin{figure}
  \centering
  \includegraphics[width=0.7\columnwidth]{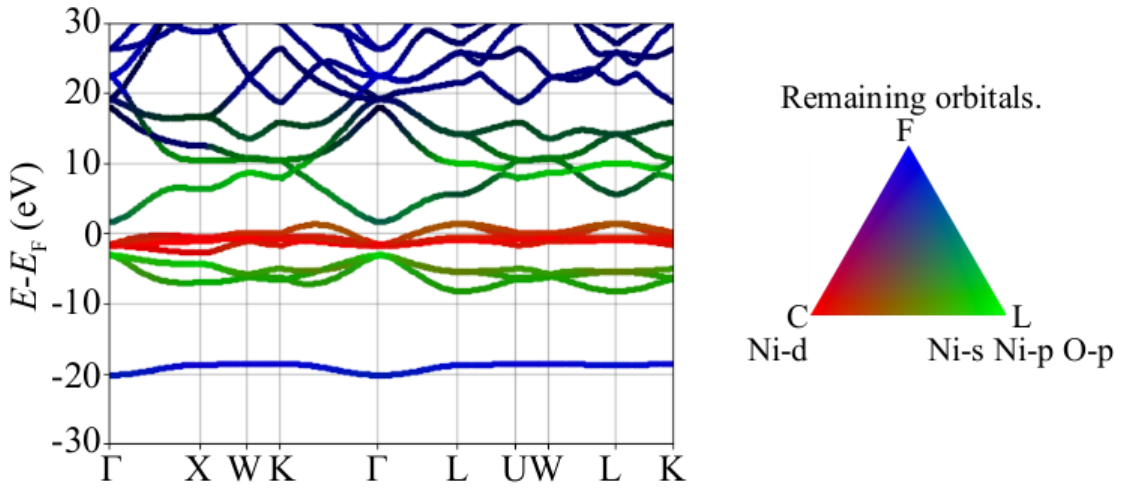}
  \caption{Three distinct Hilbert spaces (C, L, and F) and their projection onto NiO DFT eigenstates.
  }
  \label{Fig_nio_Hilbert}
\end{figure}

The low-energy space is a subspace of the full space, and a parent vector space of the correlated space. We construct this Hilbert space (L) for the crystal momentum ($\mathbf{k}$) interpolation of $G^{GW}$ and $W^{GW}$. These are two quantities obtained from \textit{ab initio} GW calculations. Due to the heavy numerical cost, it is not feasible to acquire them at an arbitrary crystal momentum resolution. Therefore, we take advantage of Wannier interpolation to interpolate these two quantities under the constraint of a restricted Hilbert space. Maximally-localized Wannier functions are used as the localized basis sets for fermionic quantities. Typically, they are constructed to span in the low-energy window (typically chosen in $E_F\pm 10eV$ as a frozen energy window). The basis functions for the bosonic quantities are constructed by calculating orthonormalized product basis functions of the low-energy fermionic basis functions. We will use lower-case Latin alphabet to label the basis functions in the low-energy space, e.g. $|\varphi_i^{f/b}\rangle$. After the basis function construction, we project $G^{GW}$ and $W^{GW}$ into the low-energy space and then interpolate them in the momentum space.

Figure \ref{Fig_nio_Hilbert} displays the three distinct Hilbert spaces (C, L, and F) and their projection onto NiO DFT eigenstates. For NiO, the correlated orbitals consist of Ni-d orbitals. The low-energy itinerant orbitals, found in the low-energy subspace but not the correlated subspace, include Ni-s, Ni-p, and O-p orbitals. The high-energy itinerant states are those in the full Hilbert space, but not in the low-energy subspace. For demonstration, we computed three quantities for each DFT eigenstate ($n\mathbf{k}\rangle$) and represented them in a band structure plot using RGB colors: $\sum_{M\in C}|\langle \varphi_M|n\mathbf{k}\rangle|^2$ (red), $\sum_{M \not\in C, M \in L}|\langle \varphi_M|n\mathbf{k}\rangle|^2$ (green), and $\sum_{M \not\in L, M \in F}|\langle \varphi_M|n\mathbf{k}\rangle|^2$ (blue). Orbitals near the Fermi energy are primarily correlated orbitals, whereas in the 5-10 eV excitation range, they consist mainly of low-energy itinerant orbitals. For excitations above 10 eV, the orbitals are predominantly high-energy itinerant states.

We also interpolate the Green's function, $G^{GW}$, and the screened Coulomb interaction, $W^{GW}$, onto a finer Matsubara frequency grid. The simulation temperature determines the frequency grid for both $G^{GW}$ and $W^{GW}$. For the \textit{ab initio} GW calculation, we employ the LqsgwFlapw package; however, GW calculations at temperatures lower than 1000K can be computationally costly. As an alternative, we approximate $G^{GW}$ and $W^{GW}$ at temperatures lower than 1000K via interpolation of the quantities obtained at 1000K.
For the frequency interpolation of Green's function, we are assuming that $\overline{G}_{GW}^{-1}(\mathbf{k}^f,\omega=0)=\left(\overline{G}_{GW}^{-1}(\mathbf{k}^f,i\omega_1^c)+\left(\overline{G}_{GW}^{-1}(\mathbf{k}^f,i\omega_1^c)\right)^\dagger\right)/2$, where $i\omega_{n=1}^c$ is the lowest Matubara frequency in the coarse frequency grid.
This is based on the fact that $\overline{G}_{GW}^{-1}$ is determined by the GW self-energy diagram and thus expected to exhibit smooth behavior near zero frequency. This has been confirmed in the cases of NiO and SrVO$_3$.

For a function (A) in the low-energy space and correlated space, we utilize $\overline{A}$ and $\widetilde{A}$ notations, respectively. To exemplify, $A$ represents the function $A$ in the full space, which is calculated on a coarse crystal momentum  and frequency grid. $\overline{A}$ denotes the same function in the low-energy space, calculated on a fine crystal momentum  and frequency grid using interpolation techniques. Lastly, $\widetilde{A}$ is denoting the function in the correlated space, calculated on a fine frequency grid.

For the projection of a parent space onto a subspace, we employ the notation $\hat{P}$, while $\hat{E}$ is used for the embedding of a subspace quantity. Subscripts on the $\hat{E}$ and $\hat{P}$ denote the initial and destination vector spaces, and $F$, $L$, and $C$ are used to represent full space, low-energy space, and correlated space, respectively. Additionally, superscripts ``f'' and ``b'' signify ``Fermionic'' and ``Bosonic''. To illustrate, $\hat{P}_{F\to L}^f(A)$ denotes the fermionic projection of $A$ from the full space to the low-energy space.

\begin{figure}
  \centering
  \includegraphics[width=0.9\columnwidth]{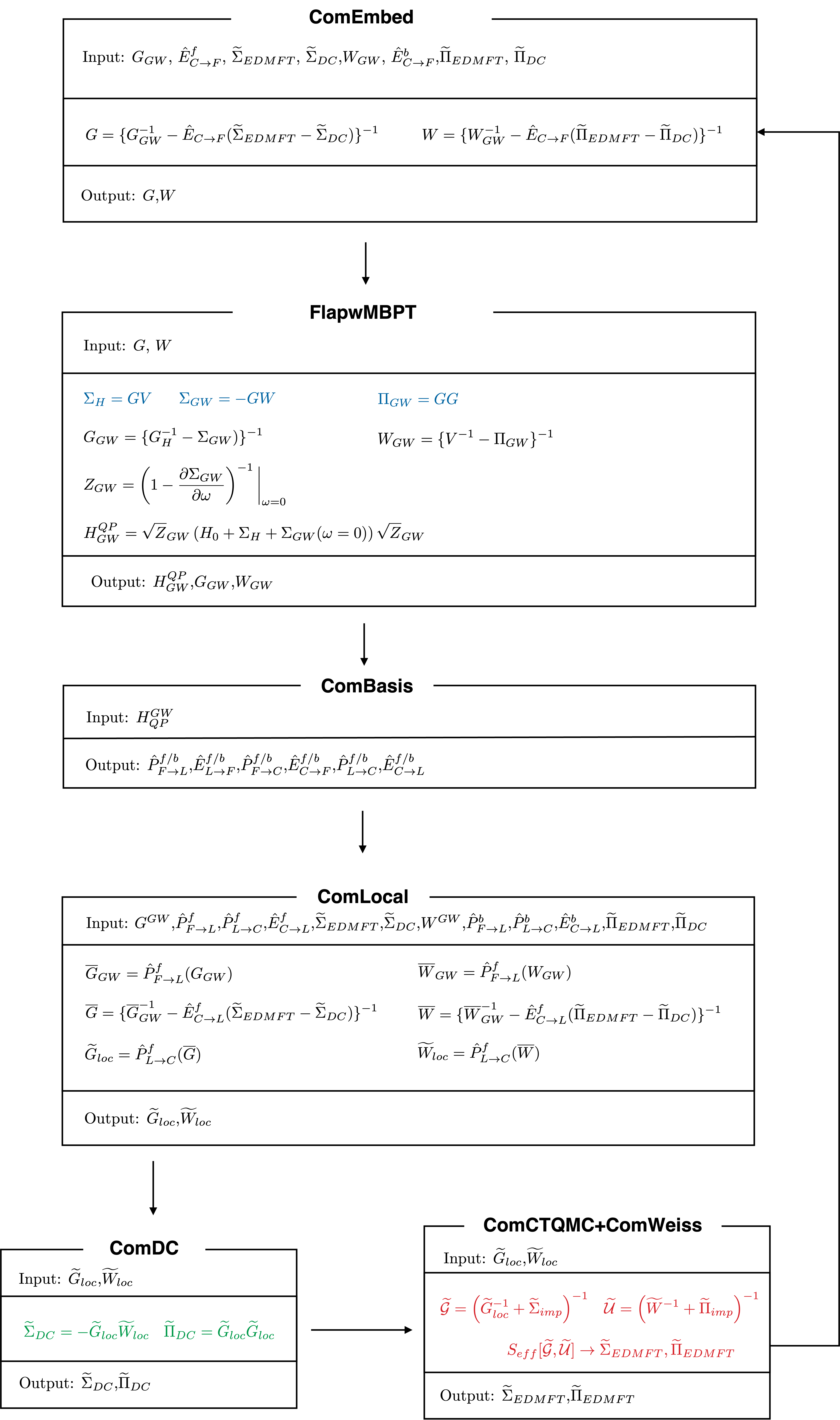}
  \caption{ComDMFT v2.0 components and the flowchart for full GW+EDMFT calculation. Inputs, outputs and important equations each component solves are also shown. Colored equations show GW (blue), EDMFT (red), and double counting (green) calculations shown in Fig. \ref{fig_gwedmft_three_scf}
  }
  \label{flowchart}
\end{figure}

Fig. \ref{flowchart} provides the complete GW+EDMFT flowchart. To begin, LqsgwFlapw computes the GW Green's function ($G_{GW}$) and GW screened Coulomb interaction ($W_{GW}$). Subsequently, ComBasis constructs localized orbitals $\varphi_i^f(\mathbf{r})$ and $\varphi_\alpha^f(\mathbf{r})$ as well as bosonic basis functions $\varphi_i^b(\mathbf{r})$ and $\varphi_\alpha^b(\mathbf{r})$. These basis functions in the low-energy space and correlated space facilitate fermionic projection ($\hat{P}^f$),  fermionic embedding ($\hat{E}^f$), bosonic projection ($\hat{P}^b$) and bosonic embedding ($\hat{E}^b$). ComLocal then calculates the local Green's function ($\widetilde{G}_{loc}$) and local screened Coulomb interaction $\widetilde{W}_{loc}$ under the GW+EDMFT approximation. Following this, ComDC obtains the double-counted self-energy ($\Sigma_{DC}$) and double-counting polarizability ($\Pi_{DC}$) within the local GW approximation. ComCTQMC and ComWeiss together evaluate the fermionic and bosonic Weiss fields, and compute the EDMFT self-energy ($\Sigma_{EDMFT}$) and EDMFT polarizability ($\Pi_{EDMFT}$). Finally, ComEmbed embeds the EDMFT and double-counting objects into $G_{GW}$ and $W_{GW}$ in the full space. This loop is iteratively solved until a converged solution is achieved. 

\subsection{LqsgwFlapw}
\subsubsection{full GW calculation}
We conducted a \textit{first-principles} study to obtain the GW Green's Function, $G_{GW}$, and the GW screened Coulomb Interaction, $W_{GW}$, using LqsgwFlapw \cite{Kutepov_Kotliar-LinearizedSelfconsistent-ComputerPhysicsCommunications-2017}. To this end, we modified LqsgwFlapw to calculate the GW self-energy, $\Sigma_{GW}$ and GW polarizability, $\Pi_{GW}$, from $G$ and $W$ within the framework of GW+EDMFT. 

\begin{equation}
  \begin{split}
    \Sigma_{GW}(\mathbf{k}^{c}, i\omega_n^{c})=&-\sum_{\mathbf{R}^{c}}\int_0^{\beta^{c}} d\tau G(\mathbf{R}^{c}, \tau)\circ W(-\mathbf{R}^{c}, -\tau) e^{-i(\mathbf{k}^{c}\cdot \mathbf{R}^{c}-\omega_n^{c}\tau)}\\
    P_{GW}(\mathbf{k}^{c}, i\nu_n^{c})=&\sum_{\mathbf{R}^{c}}\int_0^{\beta^{c}} d\tau G(\mathbf{R}^{c}, \tau)\circ G(-\mathbf{R}^{c}, -\tau) e^{-i(\mathbf{k}^{c}\cdot \mathbf{R}^{c}-\nu_n^{c}\tau)}\\
    \label{gw self_energy}
  \end{split}
\end{equation}
Here, superscript ``c'' in $\mathbf{k}^c$, $\mathbf{R}^c$, $i\omega_n^c$, $i\nu_n^c$ and $\beta^c$ denotes ``coarse'' frequency and momentum grids, and $\circ$ denotes element-wise product in the position and spin space. Then LqsgwFlapw calculates $G_{GW}$ and $W_{GW}$ in the following way.
\begin{equation}
  \begin{split}
    G_{GW}(\mathbf{k}^{c}, i\omega_n^{c})=&\left(G_H^{-1}(\mathbf{k}^{c}, i\omega_n^{c})-\Sigma_{GW}(\mathbf{k}^{c}, i\omega_n^{c})\right)^{-1}\\
    W_{GW}(\mathbf{k}^{c}, i\omega_n^{c})=&\left(V^{-1}(\mathbf{k}^{c})-P_{GW}(\mathbf{k}^{c}, i\omega_n^{c})\right)^{-1}\\
  \end{split}
\end{equation}
Hartree energy matrix element in the position representation is calculated by using $G$ within GW+EDMFT
\begin{equation}
  \begin{split}
    \Sigma_H(\mathbf{x},\mathbf{x}',\mathbf{k}^c)=&\delta(\mathbf{x}-\mathbf{x}')\frac{1}{N_{\mathbf{k}}}\sum_{\mathbf{k_1^c}}\int d\mathbf{x}'' G(\mathbf{x}'', \mathbf{x}'',\mathbf{k}_1^c, \tau=0^-)V(\mathbf{x},\mathbf{x}'',\mathbf{k}_2^c=0).\\
  \end{split}
\end{equation}

\subsubsection{Quasiparticle Hamiltonian construction}
For the purpose of Wannier interpolation of $G^{GW}$ and $W^{GW}$ in the low-energy space, localized basis sets are required. We construct Wannier functions spanning the low-energy space (Hilbert space in the energy window of $-10\ eV < E-E_F< 10\ eV$) by using the Wannier90 package. The Wannier90 package requires the best mean-field Hamiltonian to reproduce the fully interacting GW Green's function, $G^{GW}$. By linearizing the GW self-energy $\Sigma^{GW}$ in the frequency, we construct a GW quasiparticle Hamiltonian at each iteration in the following manner.
\begin{equation}
  \begin{split}
    H_{GW}^{QP}(\mathbf{k}^c)=&\sqrt{Z_{GW}(\mathbf{k}^c)}\left(H_H(\mathbf{k}^c)+\Sigma_{GW}(\mathbf{k}^c,\omega=0)\right)\sqrt{Z_{GW}(\mathbf{k}^c)}\\
    Z_{GW}(\mathbf{k}^c)=&\left(1-\frac{\partial\Sigma_{GW}(\mathbf{k}^c,\omega)}{\partial\omega}\right)^{-1}\bigg\vert_{\omega=0}\\
  \end{split}
  \label{eq:gw_qp_hamiltonian}      
\end{equation}
Here, $H_H$ denotes Hartree Hamiltonian. By diagonalizing $H_{GW}^{QP}(\mathbf{k}^c)$, we obtain energy eigenvectors $|n\mathbf{k}^c\rangle$, where $n$ is the band index. These eigenvectors are then utilized in order to calculate the inputs for the Wannier90 package. It is important to note that $H_{GW}^{QP}$ is solely constructed in order to build Wannier functions. The comparison between $H_{GW}^{QP}$ bandstructure and the spectral function of $G^{GW}$ show excellent agreement within $E_F\pm10 eV$. Please see the supplementary materials. 

\subsection{ComBasis}

Calculating local quantities such as $\widetilde{G}_{loc}$ and $\widetilde{W}_{loc}$ presents a challenge due to the need to calculate $G_{GW}$ and $W_{GW}$ on a fine $\mathbf{k}$-point grid. The accuracy of $\widetilde{G}_{loc}$ and $\widetilde{W}_{loc}$ is dependent on the momentum resolution of the aforementioned quantities; however, obtaining them \textit{from first principles} is too computationally expensive. Wannier interpolation can be employed to interpolate $\overline{G}_{GW}(\mathbf{k}^c,i\omega_n)$ and $\overline{W}_{GW}(\mathbf{k}^c,i\omega_n)$ into those on a fine $\mathbf{k}$ grid instead.

For the Wannier interpolation of fermionic quantities in low-energy space, Maximally Localized Wannier Functions (MLWFs) spanning the low-energy space are constructed. To find orbitals with maximum localizations (or minimum of the total orbital spreads), quasiparticle wavefunction information is needed for Wannier90. For this purpose, we utilized the eigenfunctions ($|\psi_{n\mathbf{k}}\rangle$) of $H_{GW}^{QP}(\mathbf{k}^c)$ as stated in Eq. \ref{eq:gw_qp_hamiltonian}. 

The other input to the Wannier90 package is trial orbitals $\langle \mathbf{r}|T_{i\mathbf{R}}\rangle$, which provide the initial condition for the total spread minimization. We employed muffin-tin orbitals of the LAPW basis set for our trial orbitals. The radial function of an initial trial orbital can be any linear combination of any muffin-tin orbitals with the same angular momentum character and that of the centered atom. ComBasis takes the linear combination which maximizes  $\sum_{n,\mathbf{k}}^{E_{froz}^{min}<E_{n,\mathbf{k}}<E_{froz}^{max}} |\langle \psi_{n\mathbf{k}}|T_{i\mathbf{k}}\rangle|^2$. Here, $E_{froz}^{min/max}$ is the lower/upper bound of the frozen energy window, and $|T_{i\mathbf{k}}\rangle$ is the Bloch sum of $|T_{i\mathbf{R}}\rangle$. 
In ComBasis, the angular momentum part of the trial orbitals is approximately defined as the cubic-harmonics $Y_{lm}$ by default when there is no spin-orbit coupling present. 
\begin{equation}
  \begin{split}
    Y_{lm}=\begin{cases}
      \frac{i}{\sqrt{2}} \left(Y_{l}^{-|m|}-(-1)^mY_{l}^{|m|}\right),& m<0\\
      Y_{l}^0,& m=0\\
      \frac{1}{\sqrt{2}} \left(Y_{l}^{-|m|}+(-1)^mY_{l}^{|m|}\right),& m>0\\      
      \end{cases}
    \label{eq:s_wso}
  \end{split},
\end{equation}
where $Y_{l}^{m}$ are the spherical harmonics. That is, a subscript $m$ corresponds to the cubic harmonics, and a superscript $m$ denotes spherical harmonics.

For a system with spin-orbit coupling, the angular part is approximately the spin-angular function ($\Omega_{l,a,m}$):
\begin{equation}
  \begin{split}
    \Omega_{l,a=\pm\frac{1}{2},m}=\sum_{s\pm 1/2}C_{a,s}^{l,m}Y_{l}^{m-s}(\hat{r})u_s.
  \end{split}
\end{equation}
Here, $u_s$ is a spinor and $C_{a,s}^{l,m}=\langle l, m-s,\frac{1}{2}, s | l+a, m \rangle$.
Then, following the procedure outlined in Sec. \ref{sec_gen_theo_framework}, we constructed orthonormalized product basis from the MLWFs ($|\psi_i^f\rangle$) spanning low-energy space and obtained a basis set for the bosonic quantities in that space.

We select a subset of orbitals with desired center positions and angular momentum characters from MLWFs and label them as correlated orbitals, which constitute the basis set of fermionic quantities in the correlated space. Subsequently, we construct an orthonormalized product basis of these correlated orbitals, thereby obtaining a basis set for the bosonic quantities in the correlated space, following the procedure detailed in Sec. \ref{sec_gen_theo_framework}.

Now, we possess all the necessary components for the projection and embedding between three distinct spaces. The projection and embedding between Full space and low-energy space can be accomplished via the following approach.
\begin{equation}
  \begin{split}
    \hat{P}_{F\to L,ij}^{f/b}(A) =\sum_{MN\in F} \langle \varphi_i^{f/b}|\varphi_M^{f/b}\rangle A_{MN} \langle \varphi_N^{f/b}|\varphi_j^{f/b}\rangle\\
    \hat{E}_{L\to F,MN}^{f/b}(\overline{A})=\sum_{ij\in L} \langle \varphi_M^{f/b}|\varphi_i^{f/b}\rangle \overline{A}_{ij} \langle \varphi_j^{f/b}|\varphi_N^{f/b}\rangle\\
  \end{split}
  \label{eq:pe_fl}              
\end{equation}
Projection and embedding between low-energy space and correlated space can be achieved in the following way.
\begin{equation}
  \begin{split}
    \hat{P}_{L\to C,\alpha\beta}^{f/b}(\overline{A})=\sum_{ij\in L} \langle \varphi_\alpha^{f/b}|\varphi_i^{f/b}\rangle \overline{A}_{ij} \langle \varphi_j^{f/b}|\varphi_\beta^{f/b}\rangle\\
    \hat{E}_{C\to L,ij}^{f/b}(\widetilde{A})=\sum_{\alpha\beta\in C} \langle \varphi_i^{f/b}|\varphi_\alpha^{f/b}\rangle \widetilde{A}_{\alpha\beta} \langle \varphi_\beta^{f/b}|\varphi_j^{f/b}\rangle\\    
  \end{split}
  \label{eq:pe_lc}              
\end{equation}
Finally, a mapping between the full space and the correlated-energy space can be constructed through the following projection and embedding.
\begin{equation}
  \begin{split}
    \hat{P}_{F\to C,\alpha\beta}^{f/b}(A) =\sum_{MN\in F} \langle \varphi_\alpha^{f/b}|\varphi_M^{f/b}\rangle A_{MN} \langle \varphi_N^{f/b}|\varphi_\beta^{f/b}\rangle\\
    \hat{E}_{C\to F,MN}^{f/b}(\widetilde{A})=\sum_{\alpha\beta\in C} \langle \varphi_M^{f/b}|\varphi_\alpha^{f/b}\rangle \widetilde{A}_{\alpha\beta} \langle \varphi_\beta^{f/b}|\varphi_N^{f/b}\rangle\\
  \end{split}
  \label{eq:pe_cf}              
\end{equation}

\subsection{ComLocal}\label{sec_comlocal}

ComLocal calculates the local Green's function ($\widetilde{G}_{loc}$) and local screened Coulomb interaction ($\widetilde{W}_{loc}$) in the following way. First, with the input of $G_{GW}(\mathbf{k}^c,i\omega_n^c)$ and  $W_{GW}(\mathbf{k}^c,i\omega_n^c)$ of LqsgwFlapw, ComLocal projects them into the low-energy space.

\begin{equation}
  \overline{G}_{GW}(\mathbf{k}^c,i\omega_n^c)=\hat{P}_{F\to L}^f\left(G_{GW}(\mathbf{k}^c,i\omega_n^c)\right)
\end{equation}
\begin{equation}
  \overline{W}_{GW}(\mathbf{k}^c,i\nu_n^c)=\hat{P}_{F\to L}^b\left(W_{GW}(\mathbf{k}^c,i\nu_n^c)\right)
\end{equation}
Next, ComLocal Wannier-interpolates them from the coarse k-grid to the fine k-grid, to obtain $\overline{G}_{GW}(\mathbf{k}^f,i\omega_n^c)$ and $\overline{W}_{GW}(\mathbf{k}^f,i\omega_n^c)$. Subsequently, the interpolation from the coarse Matsubara grid to fine-Matsubara grid is carried out to acquire $\overline{G}_{GW}(\mathbf{k}^f,i\omega_n^f)$ and $\overline{W}_{GW}(\mathbf{k}^f,i\omega_n^f)$. Then, EDMFT and double-counting quantities are embedded into $\overline{G}_{GW}(\mathbf{k}^f,i\omega_n^f)$ and  $\overline{W}_{GW}(\mathbf{k}^f,i\omega_n^f)$.
\begin{equation}
  \overline{G}(\mathbf{k}^f,i\omega_n^f)=\{\overline{G}_{GW}^{-1}(\mathbf{k}^f,i\omega_n^f)-\hat{E}_{C\to L}^f(\widetilde{\Sigma}_{EDMFT}(i\omega_n^f)-\widetilde{\Sigma}_{DC}(i\omega_n^f))\}^{-1}
\end{equation}
\begin{equation}
  \overline{W}(\mathbf{k}^f,i\omega_n^f)=\{\overline{W}_{GW}^{-1}(\mathbf{k}^f,i\omega_n^f)-\hat{E}_{C\to L}^b(\widetilde{\Pi}_{EDMFT}(i\omega_n^f)-\widetilde{\Pi}_{DC}(i\omega_n^f))\}^{-1}  
\end{equation}
Finally, we project $\overline{G}$ and $\overline{W}$ into the correlated subspace and obtain local quantities. 
\begin{equation}
\widetilde{G}_{loc}(i\omega_n^f)=\hat{P}_{L\to C}^f\left(\overline{G}(\mathbf{k}^f,i\omega_n^f)\right)
\end{equation}
\begin{equation}
\widetilde{W}_{loc}(i\omega_n^f)=\hat{P}_{L\to C}^b\left(\overline{W}(\mathbf{k}^f,i\omega_n^f)\right)
\end{equation}

\subsection{ComDC}\label{sec_comdc}
ComDC is the module to calculate the double-counted electron self-energy and polarizability. Local GW self-energy has been calculated in the following way. 
\begin{equation}
  \widetilde{\Sigma}_{DC,\alpha,\beta}(i\omega_n)=-\sum_{\gamma\delta}\int d\tau \widetilde{G}_{loc,\delta\gamma}(\tau)\widetilde{W}_{loc,\alpha\gamma\beta\delta}(\tau)e^{i \omega_n\tau}.\label{eq:dc}
\end{equation}
where $\alpha$,$\beta$,$\gamma$ and $\delta$ are composite indices for both orbital and spin (spin-angular function index) in the system without (with) spin-orbit coupling. Double-counting polarizability is calculated in the following way.
\begin{equation}
  \widetilde{P}_{\alpha\beta\gamma\delta}(i\nu_n)\allowbreak=\int d\tau
  \widetilde{G}_{\delta\beta}(\tau)\widetilde{G}_{\gamma\alpha}(-\tau)\allowbreak e^{i\nu_n\tau}.\label{eq:pi_wso}
\end{equation}

\subsection{ComWeiss}\label{sec_comweiss}
ComWeiss calculates bosonic as well as fermionic Weiss field in the following way.
\begin{equation}
\widetilde{\mathcal{G}}={\left(\widetilde{G}_{loc}^{-1}+\widetilde{\Sigma}_{EDMFT}\right)^{-1}}
\end{equation}
\begin{equation*}
\widetilde{\mathcal{U}}={\left(\widetilde{W}_{loc}^{-1}+\widetilde{\Pi}_{EDMFT}\right)^{-1}}
\end{equation*}

\subsection{ComCTQMC}
In order to solve the impurity model action as required by ComDMFT, hybridization-expansion continuous time quantum Monte Carlo (CTQMC) is employed. CTQMC is an established stochastic approach for obtaining exact numerical solutions of an impurity model, namely a small interacting system embedded in an electron bath. The EDMFT action of the impurity model, required for GW+DMFT, is 
\begin{equation}
\begin{split}
\label{equ:Action}
S = &-\sum_{\alpha\chi}  \iint_0^\beta d\tau d\tau' c^\dagger_\alpha(\tau) \widetilde{\mathcal{G}}_{\alpha\chi}^{-1}(\tau - \tau') c_\chi(\tau')   \\
&\quad\quad+\frac{1}{2}\sum_{\alpha\chi\gamma\delta}  \iint_0^\beta  d\tau d\tau' c^\dagger_\alpha(\tau) c^\dagger_\chi(\tau') \widetilde{\mathcal{U}}_{\alpha\chi\gamma\delta}(\tau - \tau') c_\gamma(\tau')c_\delta(\tau), 
\end{split}
\end{equation}
where $c^\dagger_\alpha$ creates an electron in the generalized orbital $\alpha$ (which includes both spin and orbital degrees of freedom), $\beta$ is the inverse temperature.

The present study postulates that the frequency-dependent interaction is assumed to take the following form.
\begin{equation}
\widetilde{\mathcal{U}}_{\alpha\chi\gamma\delta}(i\nu_n) = \widetilde{U}_{\alpha\chi\gamma\delta} + F^0(i\nu_n)\delta_{\alpha\delta}\delta_{\chi\gamma},
\end{equation}
We take into account only the dynamical screening of the Slater-Condon parameter $F^0$ for the sake of convenience in the numerical algorithm based on hybridization-expansion CTQMC. $\widetilde{U}_{\alpha\chi\gamma\delta}$ are frequency-independent and are approximated by their values at $i\nu_n=\infty$. The fermionic Weiss field is divided into two parts: 
\begin{equation}
\label{equ:WeissField}
\widetilde{\mathcal{G}}_{\alpha\chi}^{-1}(i\omega_n) = (i\omega_n + \mu)\delta_{\alpha\chi} - \widetilde{t}_{\alpha\chi} - \widetilde{\Delta}_{\alpha\chi}(i\omega_n),
\end{equation}
Here, $\widetilde{t}$ is the impurity-level matrix and $\widetilde{\Delta}$ is the hybridization function matrix. By solving the quantum impurity action, we obtain the self-energy ($\widetilde{\Sigma}_{EDMFT}$) and the polarizability($\widetilde{\Pi}_{EDMFT}$).

\subsection{ComEmbed}
In the final step, EDMFT and double-counting quantities were incorporated into the $G_{GW}(\mathbf{k}^c,i\omega_n^c)$ and $W_{GW}(\mathbf{k}^c,i\nu_n^c)$ of LqsgwFlapw. 
\begin{equation}
  \begin{split}
    G=\{G_{GW}^{-1}-\hat{E}_{C\to F}^f(\widetilde{\Sigma}_{EDMFT}-\widetilde{\Sigma}_{DC})\}^{-1}\\
    W=\{W_{GW}^{-1}-\hat{E}_{C\to F}(\widetilde{\Pi}_{EDMFT}-\widetilde{\Pi}_{DC})\}^{-1}\\
  \end{split}
\end{equation}
The obtained Green's function, $G(\mathbf{k}^c,i\omega_n^c)$, and the screened Coulomb interaction, $W(\mathbf{k}^c,i\nu_n^c)$, will be utilized to calculate the self-energy, $\Sigma_{GW}(\mathbf{k}^c,i\omega_n^c)$, and the polarization function, $P_{GW}(\mathbf{k}^c,i\nu_n^c)$ in eq. \ref{gw self_energy}. 

\subsection{Implementation details}
\subsubsection{Causal Optimization}

Truncations in the $\Psi$ functional of the  Luttinger-Ward (or Baym-Kadanoff) functional\cite{Baym_Kadanoff-ConservationLaws-Phys.Rev.-1961,Luttinger_Ward-GroundStateEnergy-Phys.Rev.-1960,Potthoff_Potthoff-SelfenergyfunctionalApproach-Eur.Phys.J.B-2003} provides a roadmap to build GW+EDMFT theory. Recently, it has been shown that the free energy functional can be multivalued \cite{Vucicevic_Parcollet-PracticalConsequences-Phys.Rev.B-2018,Rossi_Werner-SkeletonSeries-J.Phys.A:Math.Theor.-2015,Kim_Sacksteder-MultivaluednessLuttingerWard-Phys.Rev.B-2020,Gunnarsson_Toschi-BreakdownTraditional-Phys.Rev.Lett.-2017,Kozik_Georges-NonexistenceLuttingerWard-Phys.Rev.Lett.-2015}. Moreover, it has been shown that this multivaluedness can bring about noncausality of the Greens functions and self energies\cite{Vucicevic_Parcollet-PracticalConsequences-Phys.Rev.B-2018}. 
Noncausal Green's functions can  also arise from non-local extension of dynamical mean field theory\cite{Backes_Biermann-NonlocalCorrelation-Phys.Rev.B-2022,Biroli_Kotliar-ClusterDynamical-Phys.Rev.B-2004,Lee_Haule-DiatomicMolecule-Phys.Rev.B-2017}. 
To address this problem, we used the causal optimization method\cite{Han_Choi-CausalOptimization-Phys.Rev.B-2021} which replaces
non-causal objects with best-possible causal objects. In this implementation, we used the method to find the closest causal $\Pi_{EDMFT}$, $\mathcal{U}$, and $W_{loc}$, if
these quantities are non-causal.

\subsubsection{Database architecture}
\begin{figure*}[ht]
\centering
\includegraphics[width=0.8 \textwidth]{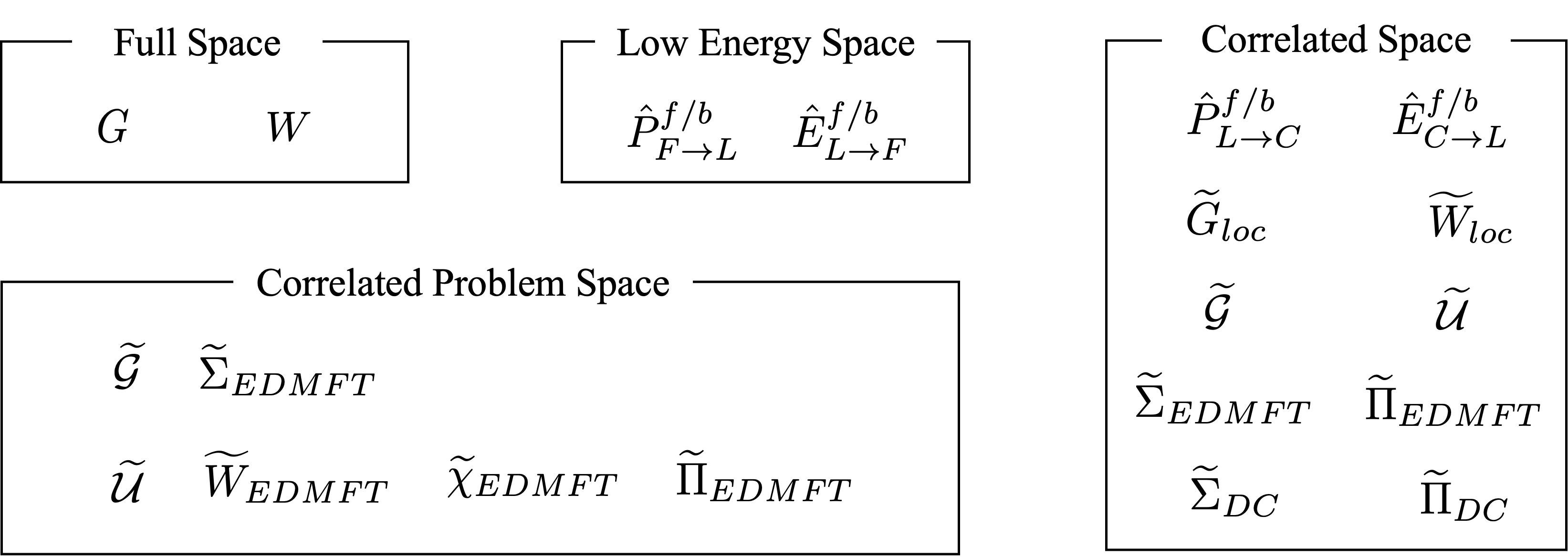}
\caption{  Grouped data structure within GW+EDMFT.}
\label{Fig_data_map}
\end{figure*}
In ComDMFT v2.0, data exchange between components is facilitated by a database. This version supports writing and reading data with the parallel HDF5 IO library. Each component stores data in a singular HDF5 file called ``global.dat''. Keys are provided to ensure the data is easily retrieved when needed. There are four HDF5 groups included in the file: ”Full Space”, ”Low energy space”, ”Correlated Space”, and ”Correlated Problem” which are denoted as different Hilbert spaces in Sec. \ref{sec_com_layout}. Data in each space is stored in its corresponding HDF5 group. Additionally, the ``Correlated Problem'' group is designed to store local quantities of each inequivalent impurity problem.
Fig. \ref{Fig_data_map} provides an overview of the groups and important data stored in the ``global.dat''. For example, $G$ and $W$ obtained from the LqsgwFlapw calculation are stored in the ``full space group'', while $\hat{P}_{F\to L}^{f/b}$ and $\hat{E}_{L\to F}^{f/b}$ are stored in the ``Low-energy space'' group.

\section{Results}\label{sec_results}
To evaluate the performance of fully self-consistent GW+EDMFT, we calculated the electronic structure of several quantum materials, 
with different degrees of correlation,  NiO and SrVO$_3$ at a temperature of 1000K.
We compared the full GW+EDMFT results with experiments, as well as with those of 
LQSGW+DMFT. We find that the intermediate quantities such as the Weiss fields are quite different in the two methods, 
but the final physical results are in rather good agreement. 

\subsection{NiO}\label{subsec_NiO}
NiO is a paradigmatic Mott insulator and it has been intensively studied 
\cite{Kunes_Vollhardt-LocalCorrelations-Phys.Rev.B-2007,Ren_Vollhardt-LDADMFT-Phys.Rev.B-2006,Karolak_Lichtenstein-DoubleCounting-JournalofElectronSpectroscopyandRelatedPhenomena-2010,Zhang_Cheng-MathrmDFTMathrmDMFT-Phys.Rev.B-2019,Kunes_Vollhardt-NiOCorrelated-Phys.Rev.Lett.-2007,Thunstrom_Eriksson-ElectronicEntanglement-Phys.Rev.Lett.-2012,Yin_Savrasov-CalculatedMomentum-Phys.Rev.Lett.-2008,Lechermann_Elsasser-InterplayChargetransfer-ArXiv190207000Cond-Mat-2019}. 
Experimentally, NiO is a paramagnetic insulator; however, traditional \textit{ab initio} electronic structure methodologies based on one-particle picture predict it to be metallic when it is not ordered magnetically.  We use NiO to illustrate the GW+EDMFT method and compare the results to LQSGW+DMFT. Both calculations treat the Ni-3$d$ orbitals as correlated orbitals, and the simulation temperature is $1000$ K.

\subsubsection{GW self-consistency}
\begin{figure*}[ht]
\centering
\includegraphics[width=0.9 \textwidth]{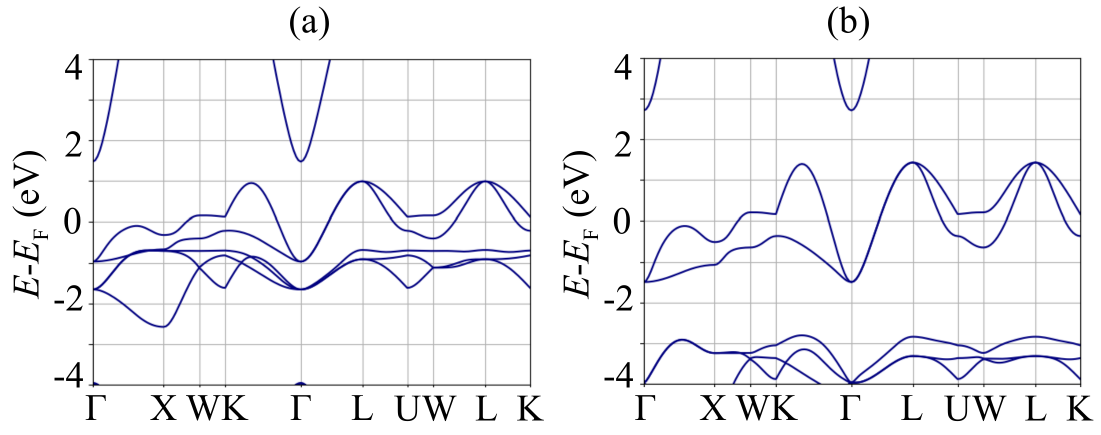}
\caption{\label{Fig_ni_gwband}
  (a) and (b) show the Wannier-interpolated GW bandstructures ($H_{GW}^{QP}$ in eq. \eqref{eq:gw_qp_hamiltonian}) of NiO at the first and last iterations of the self-consistent loop in Fig. \ref{flowchart}, respectively.}
\end{figure*}
Figure \ref{Fig_ni_gwband} (a) and (b) show the Wannier-interpolated GW bandstructures ($H_{GW}^{QP}$ in eq. \eqref{eq:gw_qp_hamiltonian}), obtained at the first and last iterations of the GW+EDMFT self-consistent loop in Figure \ref{flowchart}, respectively. The GW self-energy, as expressed in equation \ref{gw self_energy}, does not exhibit any divergence near the chemical potential. Consequently, $\Sigma_{GW}$ can be linearized safely even in this charge transfer insulator. This implies that the EDMFT self-energy, described in eq. \ref{eq:emdft_fb_self}, must have a divergence near the chemical potential to open a Mott insulating gap. Furthermore, the importance of GW self-consistency is evident when comparing figures \ref{Fig_ni_gwband} (a) and (b). The band width of two low-lying bands (largely originated from Ni-$e_g$ and O-$p$) increased by 50\%. A strong downshift of the occupied Ni-$t_{2g}$ bands is also observed. Note that this is not the GW+EDMFT bandstructure, and that the spectral function within GW+EDMFT can be obtained by embedding EDMFT and double counting self-energy to the GW Green's Function.

\subsubsection{Hybridization functions}
\begin{figure*}[ht]
\centering
\includegraphics[width=0.9 \textwidth]{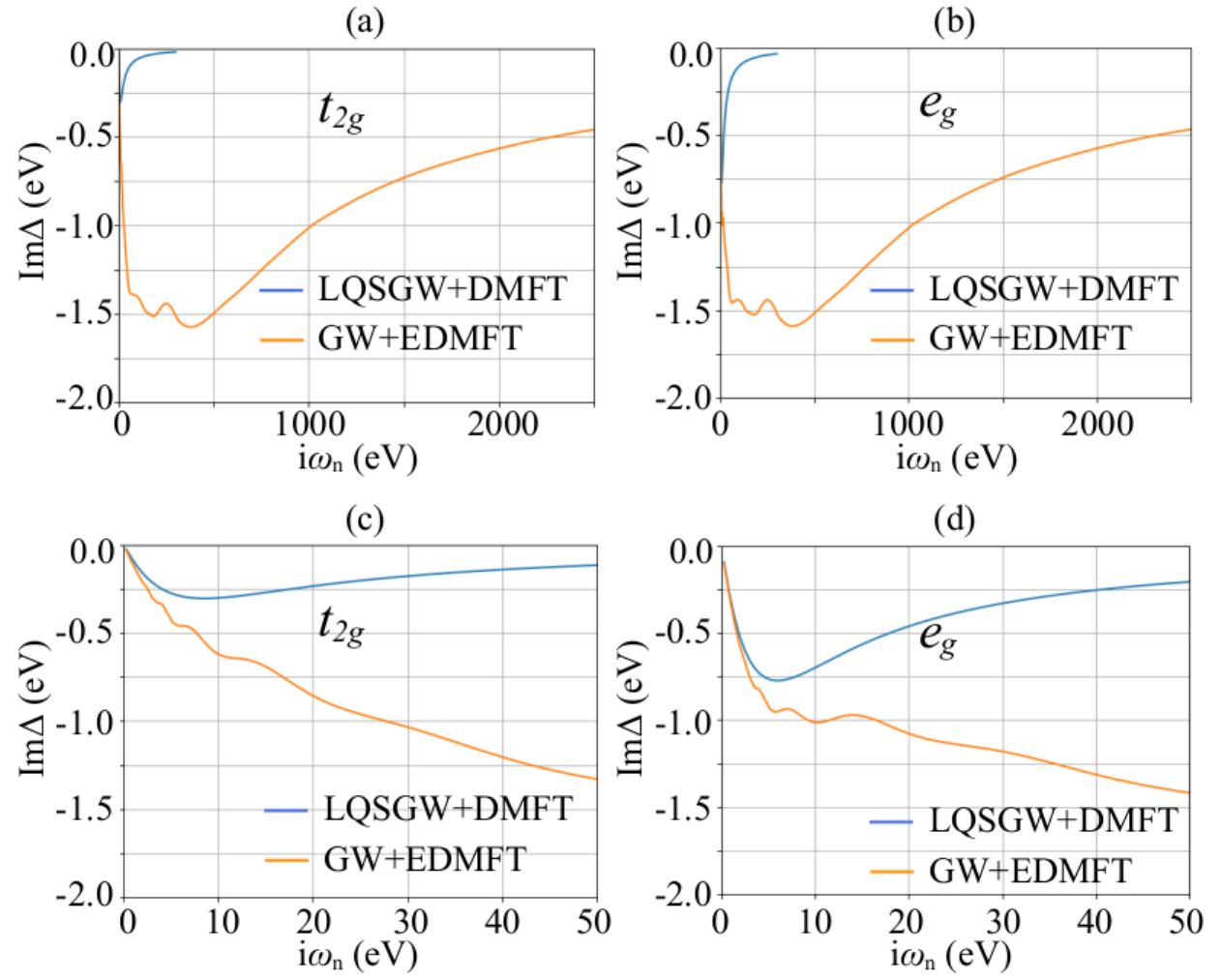}
\caption{\label{Fig_ni_hyb} Imaginary parts of hybridization functions associated with (a) Ni-$t_{2g}$ and (b) Ni-$e_{g}$ of NiO. (c) and (d) are zoom in view of (a) and (b), respectively. Orange and blue colors show the results within full GW+EDMFT and LQSGW+DMFT, respectively }

\end{figure*}
Fig. \ref{Fig_ni_hyb} illustrates the imaginary part of the hybridization functions associated with Ni-$d$ orbitals on the imaginary frequency axis. Results from both LQSGW+DMFT and GW+EDMFT are displayed in blue and orange, respectively. It is observed that Ni-$e_g$ orbitals have a stronger hybridization near the chemical potential compared to Ni-$t_{2g}$ orbitals in both methodologies. Additionally, GW+EDMFT reveals a substantially enhanced hybridization function when compared to the LQSGW+DMFT result. Within LQSGW+DMFT, it is found to be near-zero at $\sim$ 200 eV. However, a peak is present even at an energy of $\sim$ 400 eV in the GW+EDMFT hybridization function.

One possible explanation for this surprising feature of the extended tail in the hybridization function is the incoherent spectral weight introduced by ab initio GW self-energy diagrams. These diagrams introduce significant incoherence to the Green’s function, resulting in a greater deviation from the  $1/i\omega_n$  asymptotic limit in the correlated subspace. Consequently, this leads to an extended tail of the hybridization function, in contrast to DFT+DMFT or LQSGW+DMFT. In the supplementary materials, we provide a numerical demonstration of this notable feature, emphasizing the role of incoherent spectral weight from ab initio GW self-energy diagrams in extending the hybridization function’s tail.


We note that the high energy features which show up in GW+EDMFT (but no other existing method), and therefore look shocking, have little effect on the local self-energy and polarizability. Indeed, CTQMC finds identical self-energies when we dramatically adjusted these hybridization functions in order to test the consequences of these high energy features.\footnote{We forced the hybridization to smoothly transition from the low-energy to asymptotic tail regions in these tests using a spline interpolation, and then we re-ran CTQMC, to check  that the Green's functions and self-energies remained within the error bars of the solution.}


\subsubsection{Bosonic Weiss field}
\begin{figure*}[ht]
\centering
\includegraphics[width=0.9 \textwidth]{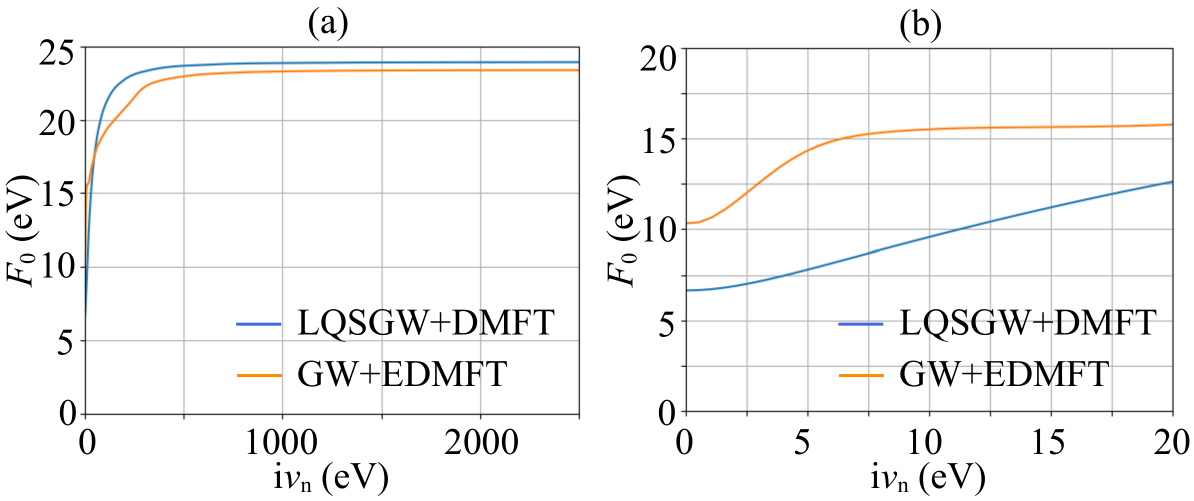}
\caption{\label{Fig_ni_bos} (a) The monopole part ($F^0$ of Slater integrals) of the bosonic Weiss field $F^{0}$ associated with Ni-$d$ orbitals in NiO. (b) zoom-in view of (a). Orange and blue colors show the results within full GW+EDMFT and LQSGW+DMFT, respectively}
\end{figure*}
Fig. \ref{Fig_ni_bos} (a) presents the monopole part ($F^0$) of the bosonic Weiss field associated with Ni-$d$ orbitals. Fig. \ref{Fig_ni_bos} (b) displays a zoomed-in view of the same figure. The static value of $F^0$ is 10 $eV$ within full GW+EDMFT, which is 40$\%$ greater than that acquired from LQSGW+DMFT. Moreover, this value further increases to 15 $eV$ at a frequency of 10 $eV$, indicating a strong frequency dependence in the low energy region. The dynamical screening captured by the Lang-Firsov approximation is suggested to be a pathway for renormalization of the hybridization function\cite{Casula_Biermann-LowEnergyModels-Phys.Rev.Lett.-2012}. Discussion on renormalization of the hybridization function due to the dynamical screening is in the following section. 



\subsubsection{EDMFT self-energies}
\begin{figure*}[ht]
\centering
\includegraphics[width=0.9 \textwidth]{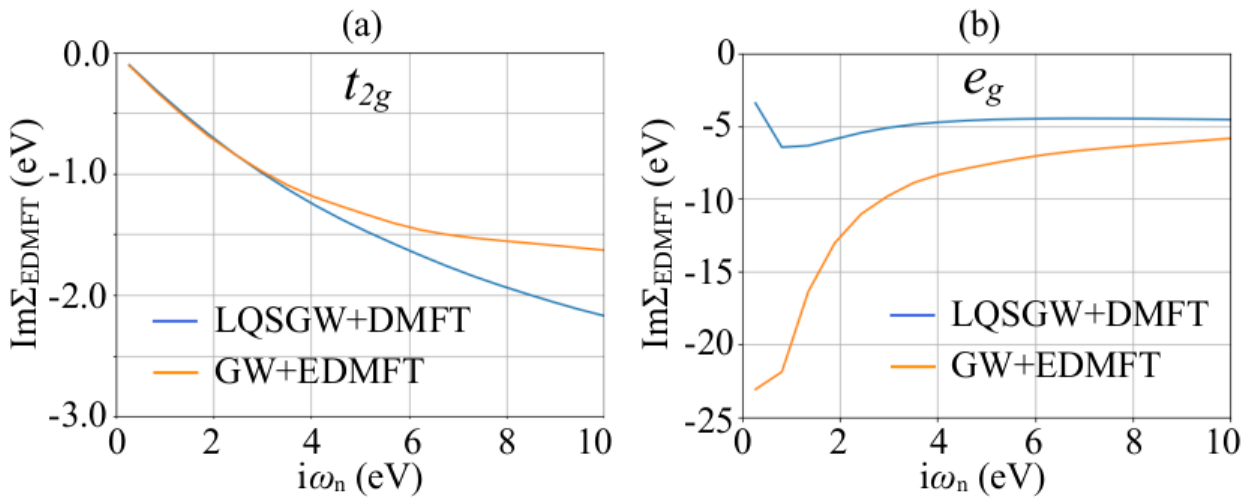}
\caption{\label{Fig_ni_sig} Imaginary parts of EDMFT self-energies associated with (a) Ni-$t_{2g}$ and (b) Ni-$e_{g}$ in NiO. Orange and blue colors show the results within full GW+EDMFT and LQSGW+DMFT, respectively}
\end{figure*}

Fig. \ref{Fig_ni_sig} presents the imaginary parts of the EDMFT self-energies for the Ni-$t_{2g}$ and Ni-$e_g$ orbitals on the imaginary frequency axis. The blue and orange lines refer to LQSGW+DMFT and GW+EDMFT, respectively. The Ni-$t_{2g}$ orbital self-energy displays a linear dependence on the imaginary frequency near the chemical potential; this is a characteristic of Fermi-liquid behavior. In contrast, the imaginary part of the Ni-$e_{g}$ self-energy reveals the onset of the divergence near the chemical potential, indicating a Mott-insulating phase. This is supported by the impurity Green's function in the static limit; on the Matsubara axis, Ni-$e_g$ impurity Green's function approches to zero values. Interestingly, the Ni-$t_{2g}$ orbital quasi-particle weights within the two different theories are found to be comparable, with values of 0.71 and 0.47 within full GW+EDMFT and LQSGW+DMFT, respectively. 
This is surprisingly similar, given the significant differences between the fermionic and bosonic Weiss fields in the two theories. This implies a compensation of the enhanced fermionic Weiss field by the enhanced bosonic Weiss fields. 

\subsubsection{Double-counting self-energies}
\begin{figure*}[ht]
  \centering
\includegraphics[width=0.9 \textwidth]{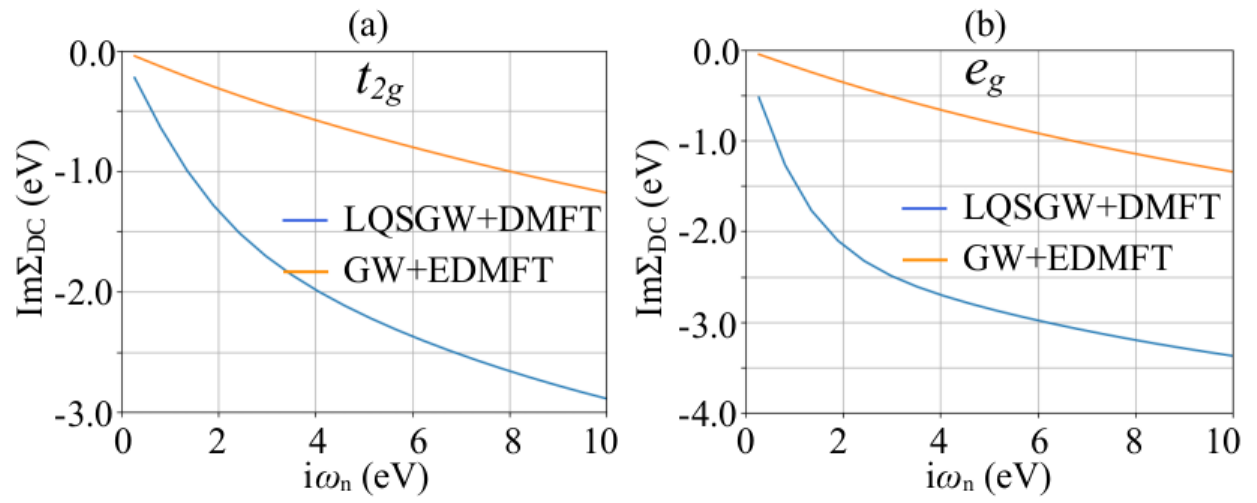}
\caption{\label{Fig_ni_dc} Imaginary parts of the double-counting self-energies associated with (a) Ni-$t_{2g}$ and (b) Ni-$e_{g}$ in NiO. Orange and blue colors show the results within full GW+EDMFT and LQSGW+DMFT, respectively}
\end{figure*}

Fig. \ref{Fig_ni_dc} displays the double-counting self-energies associated with Ni-$d$ orbitals on the imaginary frequency axis within the local-GW approximation. The results obtained from the LQSGW+DMFT and GW+EDMFT methodologies are shown in blue and orange lines, respectively. Similarly, it is observed that neither the Ni-$e_g$ nor the Ni-$t_{2g}$ orbitals exhibit any divergence in their double-counting  self-energies near zero frequency. However, there are quantitative differences in the quasiparticle-weights between the two methodologies; the Ni-$t_{2g}$ orbitals possess quasiparticle-weights of 0.88 and 0.60 in GW+EDMFT and LQSGW+DMFT, respectively, whereas for the Ni-$e_g$ orbitals, the quasiparticle-weights are 0.88 and 0.47.

\subsubsection{Spectral functions}
\begin{figure*}[ht]
\centering
\includegraphics[width=0.9 \textwidth]{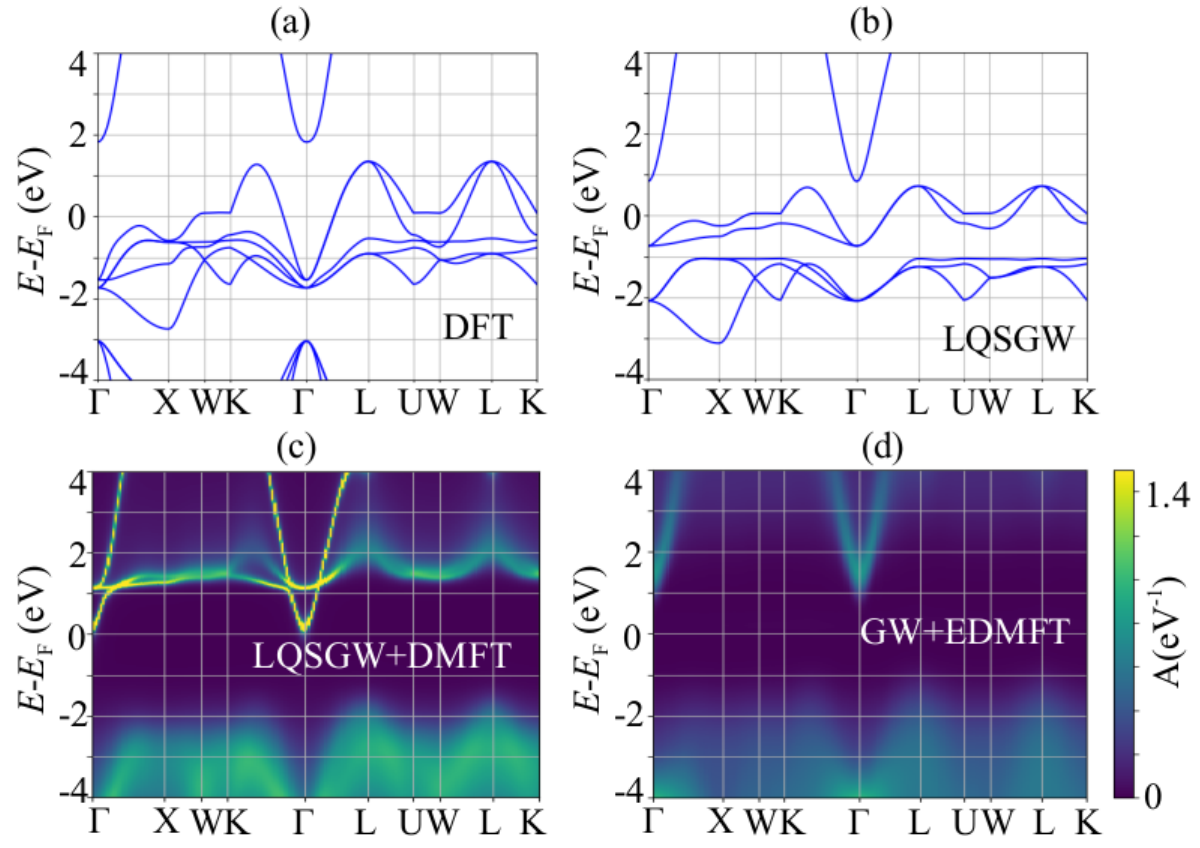}
\caption{  NiO electronic structures within (a) non spin-polarized LDA and (b) non spin-polarized LQSGW, (c) paramagnetic LQSGW+DMFT and (d) paramagnetic full GW+EDMFT, respectively}
\label{Fig_ni_band}
\end{figure*}

Fig. \ref{Fig_ni_band} presents the electronic structures of NiO without magnetic order with various  approaches:
LDA,  LQSGW, LQSGW+DMFT and  GW+EDMFT. GW+EDMFT spectral function are obtained by the the analytical continuation of the momentum-resolved Greens function on the imaginary frequency axis. Results from the non spin-polarized LDA and non spin-polarized LQSGW approaches indicate that NiO is a metal when it is not magnetically ordered, as evidenced by the presence of electronic states near the chemical potential. In contrast, GW+EDMFT and LQSGW+DMFT predict that NiO is an insulator.

\subsubsection{Projected density of states}
\begin{figure*}[ht]
\centering
\includegraphics[width=1.0 \textwidth]{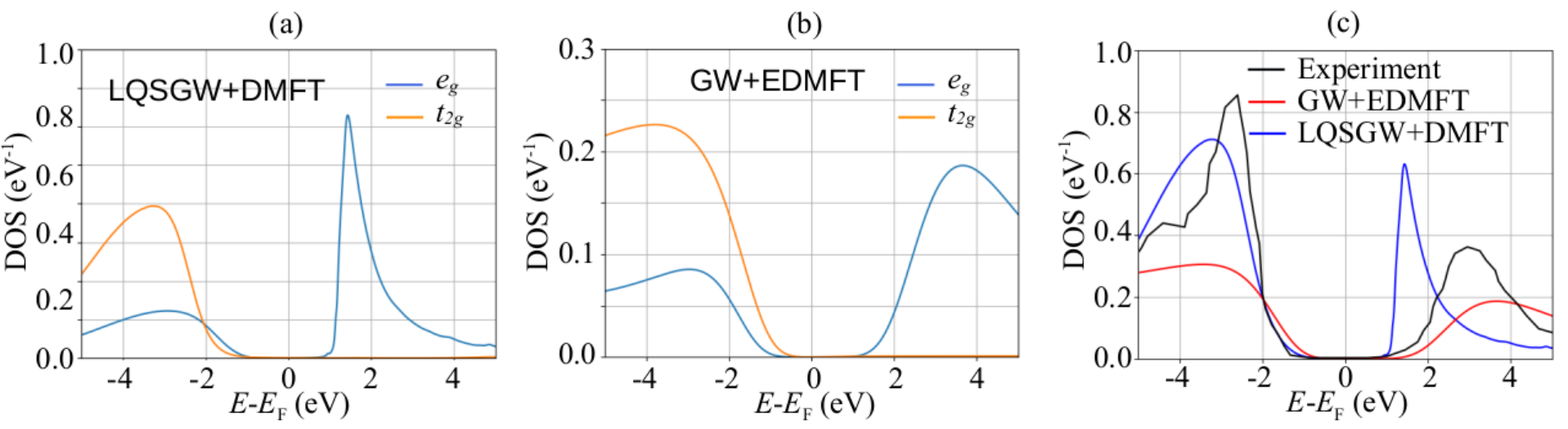}
\caption{\label{Fig_ni_dos} Projected density of states of NiO within (a) LQSGW+DMFT and (b) GW+EDMFT. (c) Comparison of the Ni-$d$ density of states with experimental result~\cite{Sawatzky_Allen-MagnitudeOrigin-Phys.Rev.Lett.-1984}.
}
\end{figure*}

Fig. \ref{Fig_ni_dos} demonstrates projected density of states (DOS) from LQSGW+DMFT and GW+EDMFT. GW+EDMFT data is obtained through analytical continuation of the orbital-resolved Greens function on the imaginary frequency axis. This data reveals that for Ni-$e_g$ orbitals, both methodologies open up a well-defined Mott insulating gap, as indicated by the strong divergence of the EDMFT self-energy. The Ni-$t_{2g}$ orbitals are fully-filled as expected.
The separation between peaks in XPS and BIS is approximately 6eV ~\cite{Sawatzky_Allen-MagnitudeOrigin-Phys.Rev.Lett.-1984}. The GW+EDMFT method slightly overestimates this value. The LQSGW+DMFT method underestimates the separation.

\subsection{SrVO$_3$}

SrVO$_3$ has been recognized to be one of the simplest correlated transition metal oxides. It is a cubic perovskite with a three V-d bands crossing the Fermi level. Photoemission and inverse photoemission spectra both exhibit satellite peaks\cite{Morikawa_Nishihara-SpectralWeight-Phys.Rev.B-1995}, making it a topical subject for DMFT methodologies\cite{Tomczak_Biermann-AsymmetryBand-Phys.Rev.B-2014,Sakuma_Aryasetiawan-ElectronicStructure-Phys.Rev.B-2013,Roekeghem_Biermann-ScreenedExchange-EPL-2014,Haule_Kotliar-CovalencyTransitionmetal-Phys.Rev.B-2014,Nilsson_Aryasetiawan-MultitierSelfconsistent-Phys.Rev.Materials-2017,Taranto_Held-ComparingQuasiparticle-Phys.Rev.B-2013,Tomczak_Biermann-CombinedGW-EPL-2012,Trimarchi_Anisimov-LDADMFT-J.Phys.:Condens.Matter-2008,Petocchi_Werner-Screeninge-Phys.Rev.Research-2020,Singh_Park-DMFTwDFTOpensource-ComputerPhysicsCommunications-2021,Chen_Marianetti-ChargeTransfer-Phys.Rev.B-2014,Nekrasov_Vollhardt-MomentumresolvedSpectral-Phys.Rev.B-2006}. In this paper, we apply GW+EDMFT to this metallic phase of SrVO$_3$ and compare the results with other methodologies. For our calculations, the V-3$d$ orbitals were treated as correlated orbitals, and the simulation temperature was set to T=1000K.

\subsubsection{GW self-consistency}
\begin{figure*}[ht]
\centering
\includegraphics[width=0.9 \textwidth]{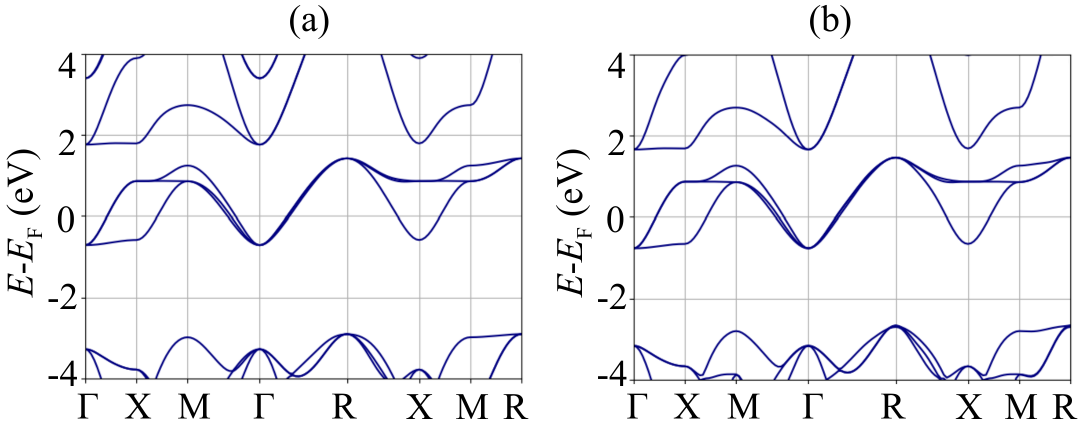}
\caption{
  (a) and (b) show the Wannier-interpolated GW bandstructures ($H_{GW}^{QP}$ in eq. \eqref{eq:gw_qp_hamiltonian}) of SrVO$_3$ at the first and last iterations of the self-consistent loop in Fig. \ref{flowchart}, respectively.
\label{Fig_sr_gwband}  
}
\end{figure*}

Fig. \ref{Fig_sr_gwband} (a) and (b) display the Wannier-interpolated GW bandstructures ($H_{GW}^{QP}$) in eq. \eqref{eq:gw_qp_hamiltonian}, obtained at the first and last iterations of the self-consistent loop in Fig. \ref{flowchart} respectively. As the GW self-energy in eq. \ref{gw self_energy} does not exhibit any divergence near the chemical potential, a linearization of the self-energy can be safely applied to construct $H_{GW}^{QP}$ in this correlated metallic phase. Notably, the GW bandstructure at the first and last iteration is almost identical, indicating that the EDMFT and double counting quantities compensate each other.

\subsubsection{Hybridization function} 
\begin{figure*}[!htb]
\centering
\includegraphics[width=0.9 \textwidth]{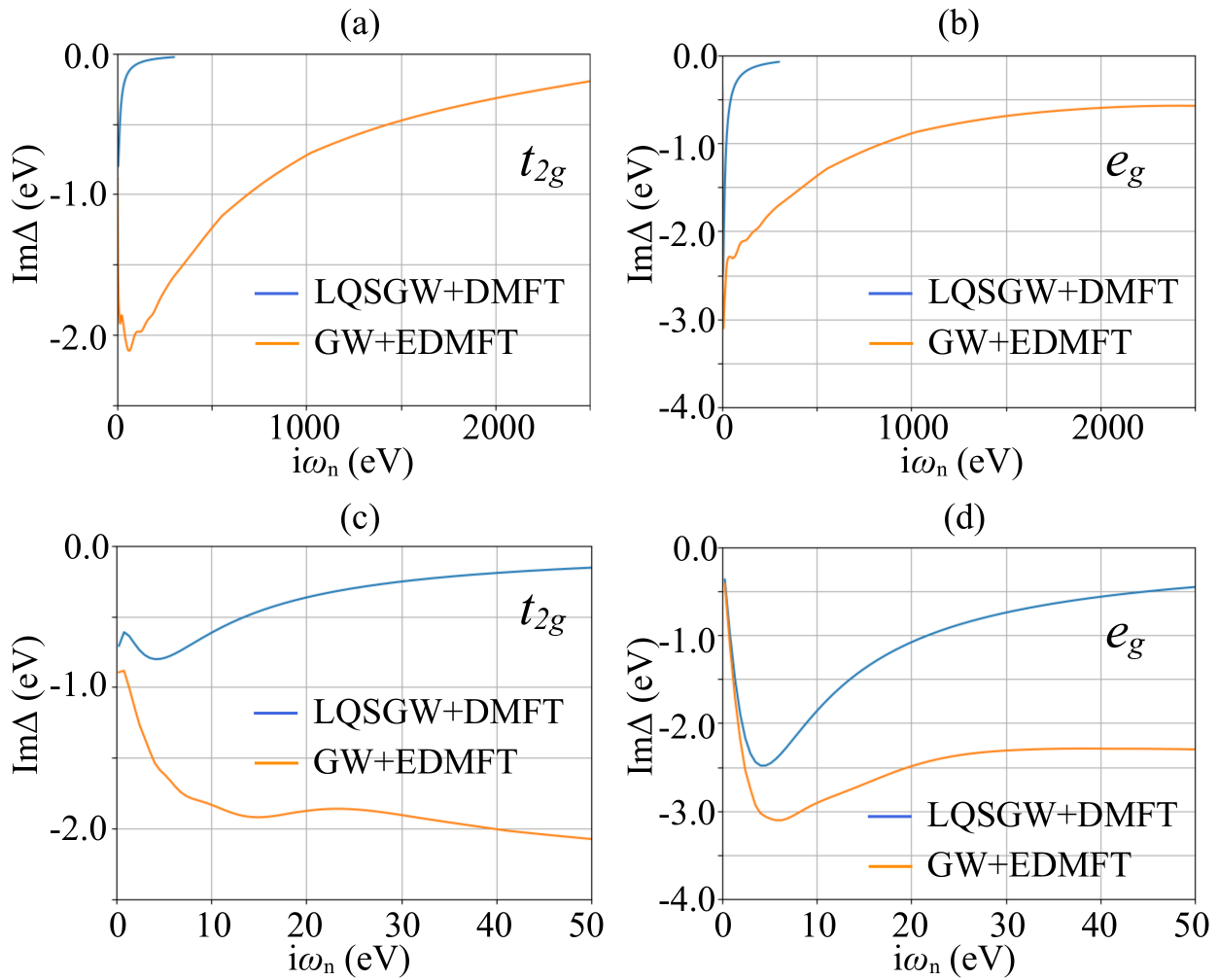}
\caption{ Imaginary parts of hybridization functions associated with (a) V-$t_{2g}$ and (b) Ni-$e_{g}$ of SrVO$_3$. (c) and (d) are zoom-in view of (a) and (b), respectively. Orange and blue colors show the results within full GW+EDMFT and LQSGW+DMFT, respectively}
\label{Fig_sr_hyb}
\end{figure*}

\begin{figure*}[!htb]
\centering
\includegraphics[width=0.9 \textwidth]{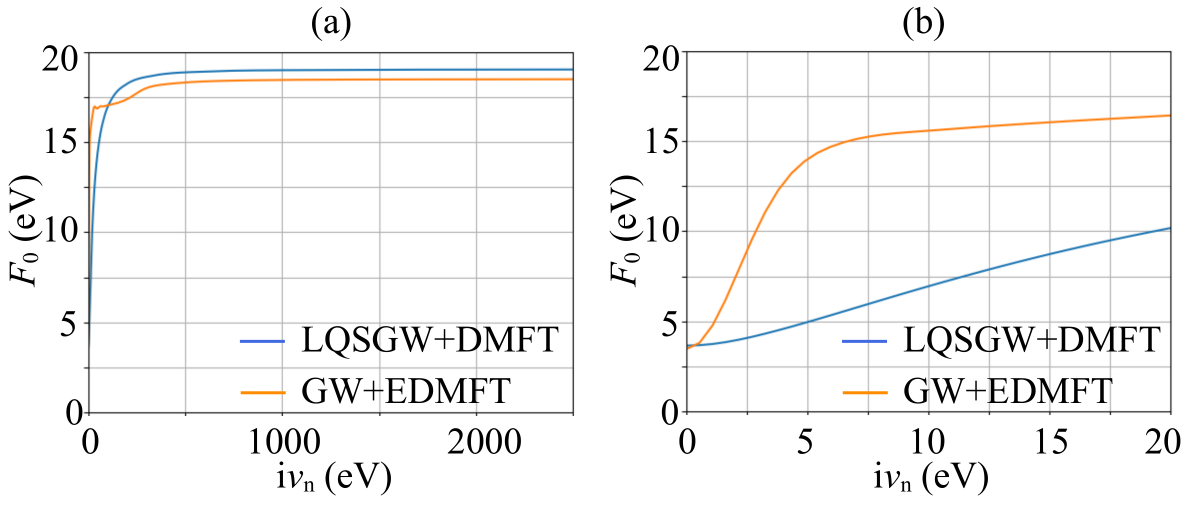}
\caption{\label{Fig_sr_bos} (a) The monopole part ($F^0$ of Slater integrals) of the bosonic Weiss field $F^{0}$ associated with V-$d$ orbitals in SrVO$_3$. (b) zoom-in view of (a). Orange and blue colors show the results within full GW+EDMFT and LQSGW+DMFT, respectively}
\end{figure*}

Fig. \ref{Fig_sr_hyb} displays the imaginary part of the hybridization functions for the V-$t_{2g}$ and V-$e_g$ orbitals on the imaginary frequency axis. Blue and orange colors in the figure correspond to the results obtained from LQSGW+DMFT and GW+EDMFT, respectively. As in NiO, the magnitude of this hybridization is very different for the same reasons discussed in Sec. \ref{subsec_NiO}. 

\subsubsection{Bosonic Weiss field}
Fig. \ref{Fig_sr_bos} (a) presents the monopole part of the bosonic Weiss field associated with V-$d$ orbitals ($F^0$ of Slater integrals). Fig. \ref{Fig_sr_bos} (b) is a zoomed-in view. The displayed blue and orange lines represent the results obtained by means of LQSGW+DMFT and GW+EDMFT, respectively. The static value of the monopole terms is 4 eV in both theories. As is the case for NiO, GW+EDMFT exhibits a stronger frequency dependence in the low-energy region, with the monopole term reaching 15 $eV$ at a frequency of 10 eV. As discussed in \ref{subsec_NiO}, this renormalizes the hybridization.

\subsubsection{EDMFT self-energies}
\begin{figure*}[ht]
\centering
\includegraphics[width=0.9 \textwidth]{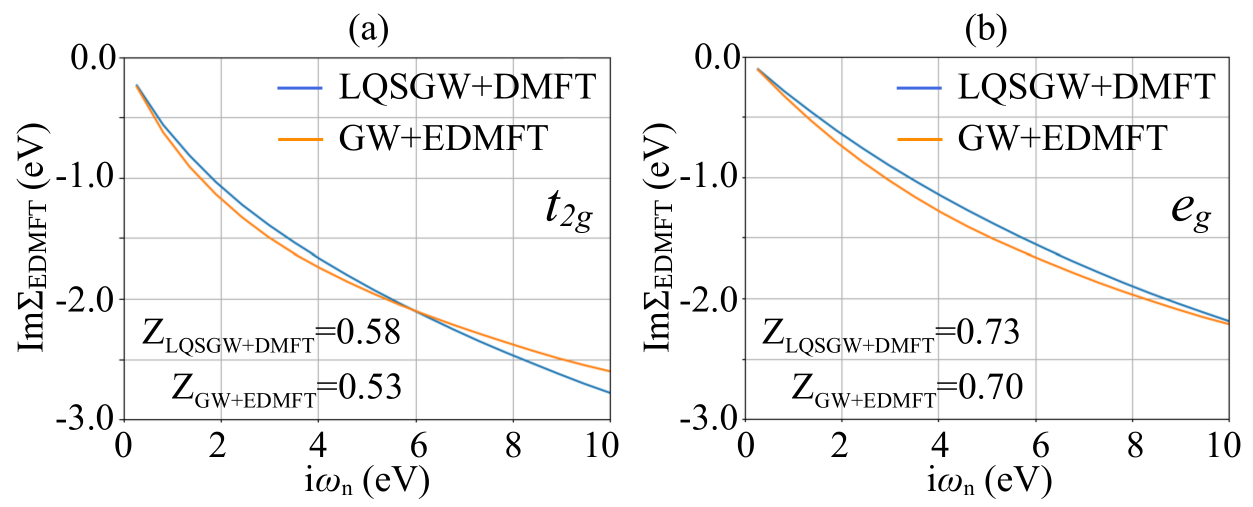}
\caption{\label{Fig_sr_sig} Imaginary parts of EDMFT self-energies associated with (a) V-$t_{2g}$ and (b) V-$e_{g}$ in SrVO$_3$. Orange and blue colors show the results within full GW+EDMFT and LQSGW+DMFT, respectively}
\end{figure*}
\begin{figure*}[ht]
  \centering
  \includegraphics[width=0.9 \textwidth]{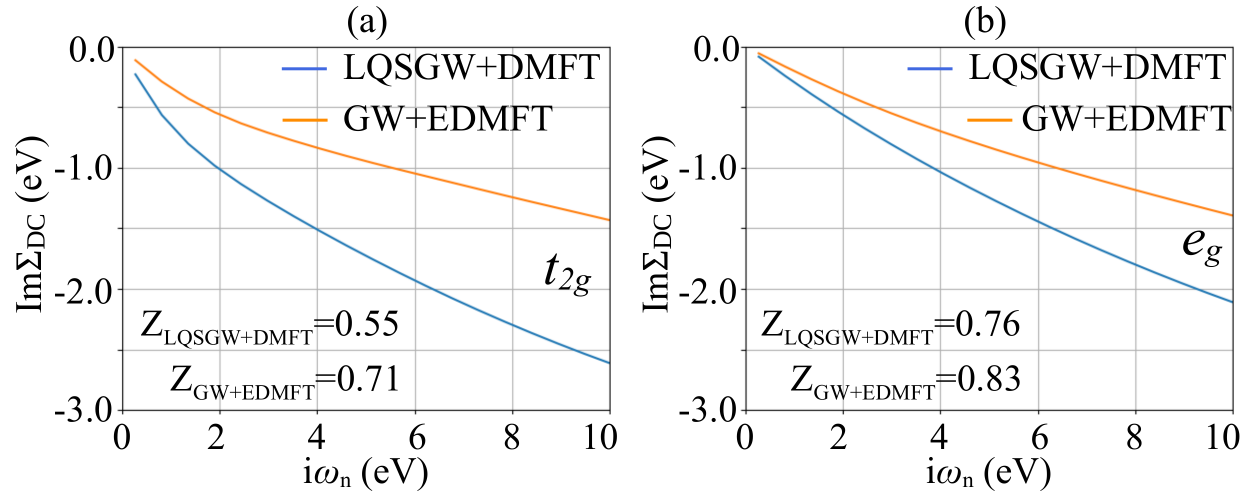}
  \caption{\label{Fig_sr_dc} Imaginary parts of the double-counting self-energies associated with (a) V-$t_{2g}$ and (b) V-$e_{g}$ in SrVO$_3$. Orange and blue colors show the results within full GW+EDMFT and LQSGW+DMFT, respectively}
\end{figure*}
Fig. \ref{Fig_sr_sig} illustrates the imaginary parts of the EDMFT self-energies for both V-$t_{2g}$ and V-$e_{g}$ orbitals as represented by the blue and orange lines, respectively, on the imaginary frequency axis. These results confirm the presence of sizable electron correlations for both orbitals. The quasiparticle weights of V-$t_{2g}$ were determined to be 0.58 and 0.53 within LQSGW+DMFT and GW+EDMFT, respectively, while the quasiparticle weights of V-$e_{g}$ were 0.73 and 0.70 within LQSGW+DMFT and GW+EDMFT, respectively. These results suggest the cancellation between the enhanced hybridization and Weiss field, as indicated by the nearly identical quasiparticle weights in both approaches. 

\subsubsection{Double-counting self-energy}
\begin{figure*}[ht]
\centering
\includegraphics[width=0.9 \textwidth]{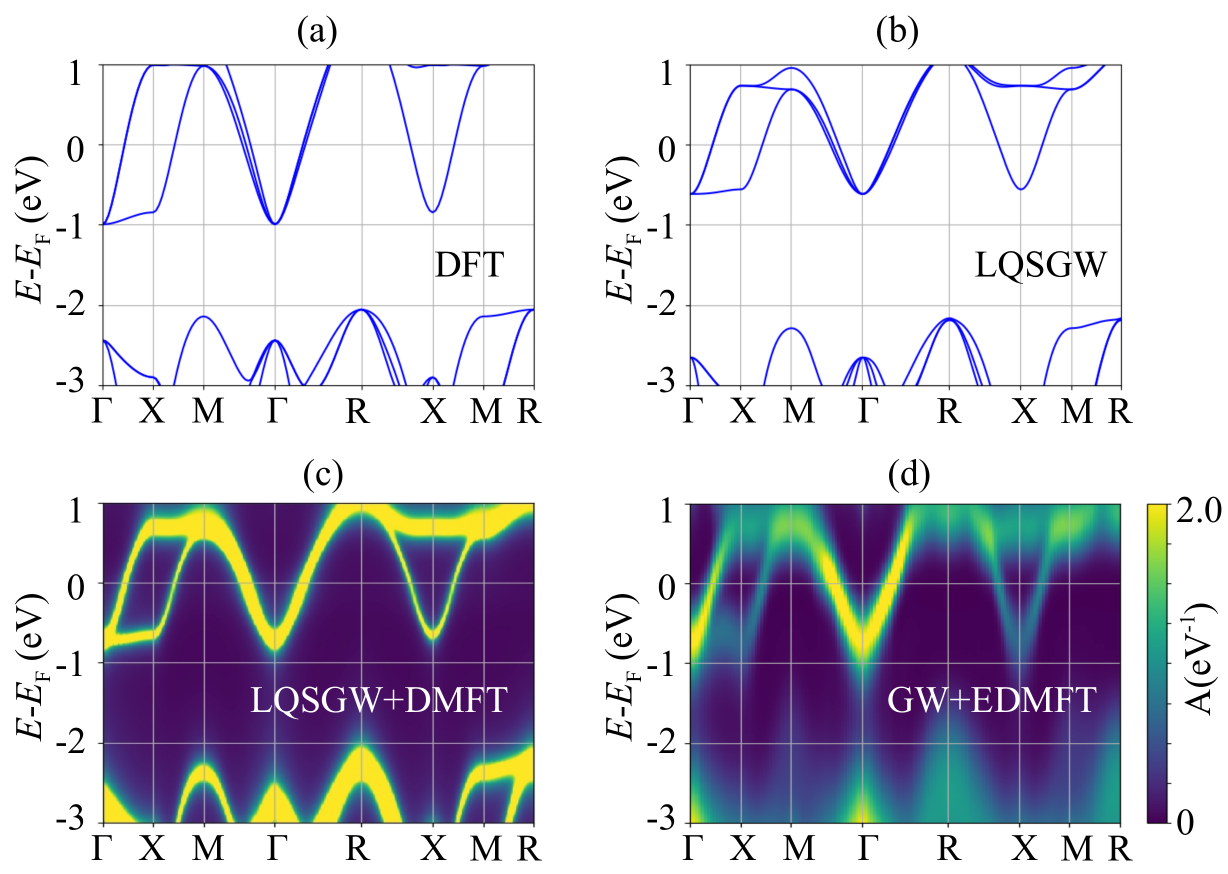}
\caption{\label{Fig_sr_band}  SrVO$_3$ electronic structures within (a) non spin-polarized LDA and (b) non spin-polarized LQSGW, (c) paramagnetic LQSGW+DMFT and (d) paramagnetic full GW+EDMFT, respectively}
\end{figure*}

Fig. \ref{Fig_sr_dc} depicts the double-counted self-energies of V-$t_{2g}$ and V-$e_{g}$ orbitals in the local-GW approximation on the imaginary frequency axis. Blue and orange lines indicate the results of LQSGW+DMFT and GW+EDMFT, respectively. It is observed that both of these approaches exhibit qualitatively similar outcomes, indicating that the V-$t_{2g}$ orbital is more correlated than V-$d_{z^2}$. Additionally, while the self-energies provided by LQSGW+DMFT and GW+EDMFT are almost identical, the quasiparticle weight in the local GW is slightly greater than that of EDMFT in GW+EDMFT.

\subsubsection{Spectral functions}
Fig. \ref{Fig_sr_band} displays the electronic structures of SrVO$_3$ computed with various \textit{ab initio} approaches: non spin-polarized Local Density Approximation (LDA), non spin-polarized LQSGW method, LQSGW+DMFT, and full GW+EDMFT. Momentum-resolved spectral function data on the real frequency axis is derived from the analytical continuation of the momentum-resolved Greens function on the imaginary frequency axis. All approaches lead to qualitatively similar band structures in the vicinity of the Fermi level. Three bands cross the Fermi level and are distinct from the other bands. The hole excitation energies at the $\Gamma$ point obtained from LDA, LQSGW, LQSGW+DMFT and GW+EDMFT are 1.0, 0.5, 0.7, and 0.7 eV, respectively, compared to the experimental value of approximately 0.5 eV\cite{Backes_Santander-Syro-HubbardBand-Phys.Rev.B-2016,Yoshida_Fujimori-CorrelatedElectronic-JournalofElectronSpectroscopyandRelatedPhenomena-2016,Mitsuhashi_Kumigashira-Influencek-Phys.Rev.B-2016,Yoshida_Fujimori-CorrelatedElectronic-JournalofElectronSpectroscopyandRelatedPhenomena-2016}.

\subsubsection{Projected density of states}
\begin{figure*}[ht]
\centering
\includegraphics[width=0.9 \textwidth]{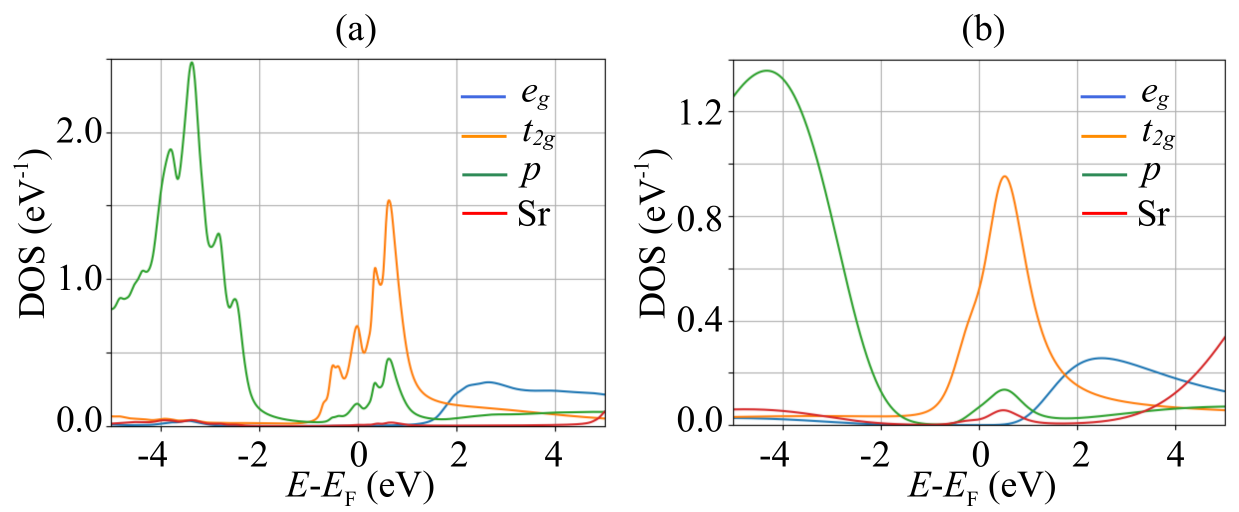}
\caption{\label{Fig_sr_dos} Projected density of states of SrVO$_3$ within (a) LQSGW+DMFT and (b) GW+EDMFT.}
\end{figure*}

Fig. \ref{Fig_sr_dos} illustrates the projected density of states of both LQSGW+DMFT and GW+EDMFT. Results from both theories present similar features, with three peak structures located at -4 eV, 0 eV, and 2 eV matching well with each other. With both theories, we did not observe strong satellite peaks on either side of the Fermi level, which agrees with the recent claim that defect states are the source of the observed satellite peak at -1.5 eV\cite{Backes_Santander-Syro-HubbardBand-Phys.Rev.B-2016}. Additionally, only a V-$e_g$ peak located at 2 eV was observed above the Fermi level.

\section{Conclusions}
We presented the implementation of fully self-consistent \textit{ab initio} GW+EDMFT for the electronic structure of correlated quantum materials in ComDMFT. This implementation enables the electronic structure calculation of quantum materials within weak, intermediate, and strong electron correlation without prior knowledge of the degree of electron correlation. We tested our implementation by calculating the electronic structures of NiO and SrVO$_3$ and obtained qualitatively similar results to LQSGW+DMFT. 

\section{Acknowledgments}
This paper was supported by the U.S. Department of Energy, Office of Science, Basic Energy Sciences as a part of the Computational Materials Science Program. S.C. was supported by a KIAS individual Grant (No. CG090601) at Korea Institute for Advanced Study. M.H. was supported by a KIAS individual Grant (No. CG091301) at Korea Institute for Advanced Study. This research used resources of the National Energy Research Scientific Computing Cen- ter (NERSC), a U.S. Department of Energy Office of Science User Facility operated under Contract No. DE-AC02-05CH11231.





\bibliographystyle{elsarticle-num}
\bibliography{gwedmft_cpc}







\end{document}